\documentclass[format=acmsmall]{acmart}

\usepackage{booktabs} 
\usepackage{subcaption}

\usepackage[ruled]{algorithm2e} 

\SetAlFnt{\small}
\SetAlCapFnt{\small}
\SetAlCapNameFnt{\small}
\SetAlCapHSkip{0pt}
\IncMargin{-\parindent}

\newlength{\myitemsep}
\newcommand{\todo}[1]{}
\renewcommand{\todo}[1]{{\color{red}[[TODO: {#1}]]}}
\newcommand{\lett}[1]{({\textbf{#1}})}

\newcommand{\Gtrue}{\ensuremath{G_\mathrm{true}}}
\newcommand{\Gsampled}{\ensuremath{G_\mathrm{sampled}}}
\newcommand{\Vtrue}{\ensuremath{V_\mathrm{true}}}
\newcommand{\Vsampled}{\ensuremath{V_\mathrm{sampled}}}

\newcommand{\Btrue}{\ensuremath{B_\mathrm{true}}}
\newcommand{\Bsampled}{\ensuremath{B_\mathrm{sampled}}}
\newcommand{\Ntrue}{\ensuremath{N_\mathrm{true}}}
\newcommand{\Nsampled}{\ensuremath{N_\mathrm{sampled}}}
\newcommand{\Mtrue}{\ensuremath{M_\mathrm{true}}}
\newcommand{\Msampled}{\ensuremath{M_\mathrm{sampled}}}

\def\detnum{determination number CHRBSS: 15-039}

\received{April 2018}
\received[revised]{July 2018}
\received[accepted]{September 2018}

\begin{document}
\title[Efficient Crowd Exploration of Large Networks]{Efficient Crowd Exploration of Large Networks: The Case of Causal Attribution}

\author{Daniel Berenberg}
\affiliation{%
  \institution{University of Vermont}
  \department{Computer Science}
  \city{Burlington}
  \state{VT}
  \postcode{05405}
  \country{USA}}
\email{daniel.berenberg@uvm.edu}
\author{James P.~Bagrow}
\affiliation{%
  \institution{University of Vermont}
  \department{Mathematics \& Statistics and Vermont Complex Systems Center}
  \city{Burlington}
  \state{VT}
  \postcode{05405}
  \country{USA}
}
\email{james.bagrow@uvm.edu}

\begin{abstract}
Accurately and efficiently crowdsourcing complex, open-ended tasks can be difficult, as crowd participants tend to favor short, repetitive ``microtasks''.
We study the crowdsourcing of large networks where the crowd provides the network topology via microtasks.
Crowds can explore many types of social and information networks, but we focus on the network of causal attributions, an important network that signifies cause-and-effect relationships.
We conduct experiments on Amazon Mechanical Turk (AMT) testing how workers propose and validate individual causal relationships and introduce a method for independent crowd workers to explore large networks.
The core of the method, Iterative Pathway Refinement, is a theoretically-principled mechanism for efficient exploration via microtasks.
We evaluate the method using synthetic networks and apply it on AMT to extract a large-scale causal attribution network, then investigate the structure of this network as well as the activity patterns and efficiency of the workers who constructed this network.
Worker interactions reveal important characteristics of causal perception and the network data they generate can improve our understanding of causality and causal inference.
\end{abstract}

%
%

%
%


\setcopyright{acmlicensed}
\acmJournal{PACMHCI}
\acmYear{2018} \acmVolume{2} \acmNumber{CSCW} \acmArticle{24} \acmMonth{11} \acmPrice{15.00}\acmDOI{10.1145/3274293}

\maketitle

\renewcommand{\shortauthors}{D.\ Berenberg \& J.\ P.\ Bagrow}


\section{Introduction}

Crowdsourcing has emerged as a powerful technique for gathering data that are otherwise inaccessible, either computationally or logistically~\cite{brabham2008crowdsourcing,howe2006rise}.
These data may be training data for machine learning algorithms, survey data, or the results of behavioral experiments~\cite{snow2008cheap,kittur2008crowdsourcing}.
While data gathering is a key use of crowdsourcing, crowd participants are also uniquely capable of ingenuity and creativity, and the most powerful applications of crowdsourcing exploit this to provide crowdsourcers with novel ideas and out-of-the-box thinking~\cite{kittur2010crowdsourcing,aragon_collaborative_2011,siangliulue2015toward}.

When applying crowdsourcing, there is often an antagonism between the complexity of the task the crowdsourcer wishes to accomplish and the preference of crowd workers in favor of short \emph{microtasks}~\cite{kittur2011crowdforge}.
This has led to considerable research on decomposing various large-scale tasks into microtasks more suitable for the crowd~\cite{cheng2015break}.
In our case, we are interested in enabling the crowd to efficiently explore large, unknown networks by asking workers to propose some of the nodes and links within a particular network of interest. 

The objective of this work is to study how independent crowd workers can efficiently explore a large causal attribution network.
We contribute a new algorithm for crowdsourcing large networks, and perform simulations, conduct experiments and perform surveys to assess the activity patterns, efficiency, and efficacy of workers using this algorithm.
%
We focus on causal attribution networks, where workers are asked to provide directed links between causes and effects, and so we also conduct experiments on Amazon Mechanical Turk to better understand how workers attribute causes and effects.
Causal reasoning, while affected by cognitive biases, remains one of the biggest differentiators between human intelligence and machine learning methods, making this problem domain an ideal venue for crowdsourcing.
Other types of networks, such as knowledge graphs, concept maps, or social networks, may also be explored with crowds.

A simple way to decompose the larger network exploration task into microtasks is to ask workers to validate a single link for each task, but this provides minimal information per task and may not be efficient.
Motivated by theoretical studies of network search strategies, 
we propose and evaluate a network exploration microtask---\textbf{Iterative Pathway Refinement}---where workers create and modify \emph{pathways}, short linear paths of nodes, within the larger network. 
These pathways are easy for workers to modify, provide more information than can be gathered from a single link microtask, and the union of these pathways provides inference of the larger network being explored.


The rest of this work is organized as follows.
Section~\ref{sec:background} surveys previous research on several aspects of the problem we study, including crowdsourcing, causal attribution, and network search.
Section~\ref{sec:exp1} describes our first experiment conducted on Amazon Mechanical Turk (AMT). 
The goal of this experiment is to estimate the efficacy of crowd workers tasked with proposing and validating individual cause-and-effect relationships.
Motivated by this experiment and with the goal of maximizing crowd efficiency,
in Sec.~\ref{sec:algorithms} we introduce a set of algorithms enabling a crowd of workers to collectively explore a large network while individually participating only in microtasks.
Then, Sec.~\ref{sec:exp2} presents our second experiment implementing these algorithms on AMT to derive a large-scale causal attribution network, 
while Sec.~\ref{sec:followupsurvey} performs a followup analysis on the quality of responses generated with our new algorithms.
Section~\ref{sec:model} proposes a mathematical model for the exploration algorithm and analyzes it on known test networks, to better understand how a sampled network derived via our exploration algorithm differs from a true, underlying network.
We conclude with a discussion in Sec.~\ref{sec:discussion}, including how our work here can be generalized and can inform the crowd exploration of other types of networks.

\section{Background}
\label{sec:background}

The focus of this work can best be understood in the context of three research areas: crowdsourcing, causal attribution theory, and search in networks. 

\subsection{Crowdsourcing}

Crowdsourcing is the recruitment and application of large groups of individuals towards the generation of work~\cite{howe2006rise,brabham2008crowdsourcing,estelles2012towards}.
The crowd may be voluntary participants or paid workers, and they may or may not need to be subject-matter experts in areas relevant to the particular crowdsourced tasks. 
Common crowdsourcing tasks include labeling images, disambiguating written records, and participating in surveys and behavioral experiments.
Yet, the experience and creativity of crowd workers is one of the most unique aspects of crowdsourcing, and leveraging these assets can achieve results far beyond the confines of ``artificial artificial intelligence''~\cite{kittur2010crowdsourcing}.

The most popular online platform for recruiting paid workers is Amazon Mechanical Turk (AMT).
Large crowdsourcing tasks are generally broken down into ``microtasks'' referred to on AMT as Human Intelligence Tasks (HITs). 
Dividing a large task or set of tasks into many small microtasks, either manually or algorithmically, is one of the most effective ways to distribute complex work over a crowd~\cite{little2010turkit,kittur2011crowdforge}.
Many methods use a propose-and-vote/fix-verify/select-validate mechanism for decomposing complex tasks without harming reliability or quality~\cite{chilton_cascade:_2013,bernstein_soylent:_2015,salganik2015wiki,bagrow2018crowd}.
Monitoring and improving upon the completion times of microtasks and batches of microtasks is an important aspect of crowdsourcing, as it can improve overall efficiency and quality~\cite{kittur2008crowdsourcing,jacques2013crowdsourcing,Difallah:2015:DMC:2736277.2741685}.
Efficiency and quality are also enhanced by using statistical aggregation strategies that combine multiple worker responses to microtasks~\cite{dawid1979maximum,karger2011iterative,kruger2014axiomatic}. 

Many crowdsourcing projects study the problem of assigning workers to a fixed set of predetermined tasks~\cite{Li:2016:CHQ:2835776.2835797}, for example annotating a collection of images, but a growing body of work is considering areas where the crowd contributes new tasks to the crowdsourcer~\cite{bongard_crowdsourcing_2013,bevelander_crowdsourcing_2014,siangliulue2015toward,salganik2015wiki,teevan_supporting_2016,wagy2017crowdsourcing}.

Crowdsourcing the collection of network data is an under-explored area, particularly in the context of non-expert participants.
One study considers crowdsourcing a network of synonymous terms---essentially a thesaurus---as a testbed for algorithms to efficiently distribute workers across a growing a set of tasks~\cite{crowdSteering}.
Another interesting study is the DREAM predictive signaling network challenge, where
research teams were challenged to determine protein signaling networks from experimental data~\cite{Prillmr7,marbach2012wisdom,hill2016inferring}.
This challenge is specific to a single research area, and only experts in that area can reasonably contribute work.
In general, it remains an open question how best to use crowdsourcing to explore large networks.

\subsection{Causal attribution}

Identifying and understanding causal relationships is a crucial way humans comprehend the world around them.
Causal inference has been a major focus of philosophy, psychology, mathematics, and statistics for centuries~\cite{hume2012treatise,kant1998critique,granger1969investigating,rubin2011causal,pearl2009causality,girju2002text,kim2013mining}. While much progress has been made developing statistical tools, establishing causal relationships remains an outstanding scientific challenge.

Human understanding and perception of cause and effect relationships is complicated and influenced by language structure~\cite{kelley1967attribution,taylor1975point,brown1983psychological,hilton1990conversational} and sentiment~\cite{bohner1988triggers}. 
The famous perception experiments of Michotte \emph{et al.} have led to a long thread of experiments exploring how and why cognitive biases affect causal attribution~\cite{joynson1971michotte,scholl2000perceptual,rolfs2013visual}.
Accounting for such biases is crucial to better understand causal attribution at scale.

\subsection{Network exploration and search}

A key question within Network Science is the problem of network exploration: how can an agent with only local or partial information understand the global structure of a large network? 
Likewise, how can an agent moving within a network efficiently find a predetermined search target?
Search strategies are useful both for finding a given target and for efficiently mapping out the underlying structure of the network topology.

The ability to efficiently identify a target node in a network is a problem that has been studied since the seminal ``small-world'' work of Travers and Milgram~\cite{travers1967small}.
Many successful strategies exploiting only local information are known for spatial networks and power-law networks~\cite{kleinberg2000navigation,adamic2001search}.
One such strategy to explore a network is to preferentially seek out the highest-degree nodes or hubs, those nodes with the most connections~\cite{adamic2001search}. 
The more quickly a searcher arrives at a hub, the more avenues it has to explore the rest of the network, although this may lead to a biased view of the hubs if the network is not fully explored.
Another well-supported local strategy is to perform a \emph{self-avoiding walk} (SAW), moving randomly over the nodes of the network without returning to any previously visited nodes~\cite{adamic2001search,PhysRevE.71.016107}. 
The established success of SAW search provides a motivation and theoretical underpinning for the key phase of the crowdsourcing algorithm we introduce in this work.

\section{Experiment 1 --- single link learning}
\label{sec:exp1}

This experiment tests the ability of crowd workers to provide causal attribution information. 
Each crowdsourcing task focuses on asking workers to validate a single cause-effect pair by asking, for example, ``\emph{Does `intelligence' cause `foresight'?}''.
These candidate cause-effect term pairs ($A,B$) (`intelligent', and `foresight' in this case) were tested by workers,
who could respond with one of several multiple choice answers: (i) ``$A$ causes $B$'', (ii) ``$B$ causes $A$'', (iii) ``$A$ and $B$ are unrelated'', (iv) ``something else causes both''.

Aggregating multiple choice responses from multiple workers as they examine different term pairs allows this approach to study a larger network of causes and effects.
However, the focus of Experiment 1 is estimating the efficacy of worker's causal attribution using various benchmark datasets. In Experiment 2 (Sec.~\ref{sec:exp2}) we return to the problem of network exploration.

\subsection{Materials and methods}

To study the efficacy of worker attributions of causal relationships, we extracted candidate cause-effect term pairs from two datasets: 

\begin{description}
\item[Ground Truth data] A benchmark database of 85 established cause-effect pairs~\cite{mooij2016distinguishing}. 
These data are intended for validating causal detection algorithms.
Many pairs, though not all, tend to cover scientific domains.
Updated versions of this dataset currently provide 108 total term pairs.

\item[Word Association data] A set of 90 term pairs sampled randomly from the University of South Florida Free Association Norms dataset~\cite{nelson2004university}. 
These data were collected over several decades from more than 6,000 participants. 
Each participant was shown a stimulus word and asked to produce the first word that comes to mind that was meaningfully related or associated with the stimulus.
These associations provide a baseline for us to test; the terms in a pair may be causally related or the association may be due to other factors.
\end{description}

Further, we introduced \textbf{randomized versions} of both datasets by shuffling the terms between pairs. 
These randomized term pairs provide a base set of approximate known negatives to see what rate workers may attribute causal connections when a relationship is unlikely.
Randomizing the same terms controls for the overall domains and contexts of the data while breaking any preexisting connections, either causal or associational, between terms in a pair.
It is possible, of course, that a strong relationship between two random terms may exist, but this is less likely for the randomized data than for the original terms.
A total of 350 distinct term-pairs were developed across the four datasets.

Data were collected on Amazon Mechanical Turk.
Workers were shown a multiple choice web form for each HIT, asking them to select one of four options relating cause-effect pair ($A,B$): $A$ causes $B$, $B$ causes $A$, $A$ and $B$ are unrelated, something else causes both.
An example of this form is shown in Fig.~\ref{fig:exp1exampleSurvey}.
As instructions, workers were shown the HIT web form populated with an example cause-effect term pair (``Population growth'', ``Food consumption growth'')  before their first HIT. 
Workers were rewarded \$0.04 per response.
We aimed to collect $n=50$ responses for each of the 350 distinct term-pairs.

This research procedure has been approved by our IRB (\detnum{}).

\subsection{Results}

We gathered 17,556 responses from 726 Mechanical Turk workers. 
Workers could complete as many HITs as available, but could not respond to the same term pair more than once.
The mean and median numbers of responses per worker was $24.18$ and $3$,  respectively.

\begin{figure}
\centering
{\includegraphics[width=0.55\textwidth]{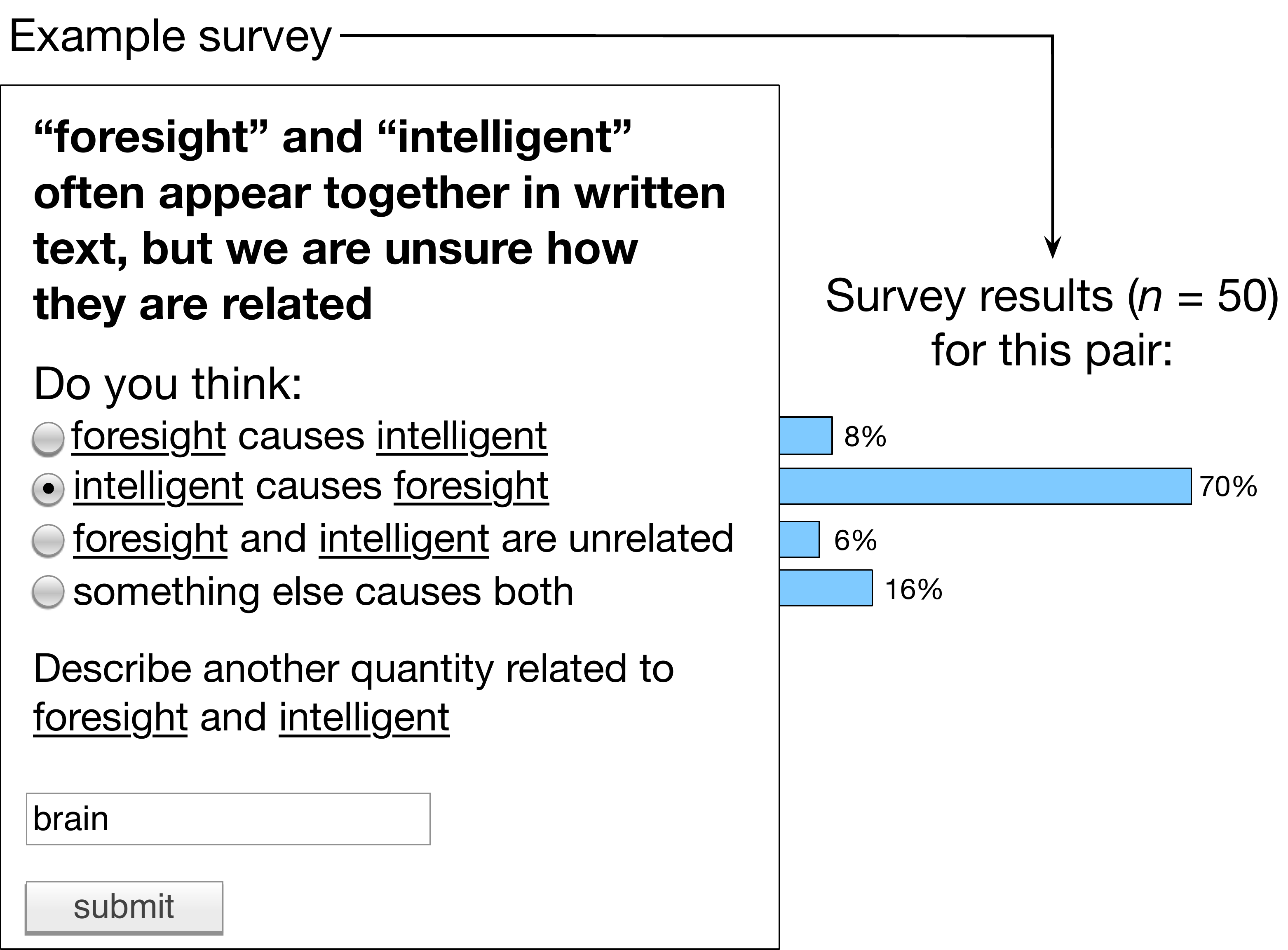}}
\caption{Example survey conducted as part of Experiment 1, in this case for the terms ``foresight'' and ``intelligent''.
The crowd reached strong consensus that intelligence causes foresight. 
\label{fig:exp1exampleSurvey}
}
\end{figure}

We aggregated the total crowd responses to each term-pair and classified the link between the terms as either `causal', `confounded', or `unrelated' based on the majority response from workers shown that pair. 
For example, the pair (``intelligent'', ``foresight'') shown in Fig.~\ref{fig:exp1exampleSurvey} had 78\% causal responses (attributing causality in either direction) so we classify it as a causal link.
In this particular term pair, the direction of causality is clearly supported by the crowd, but other pairs with a majority `causal' classification may be mixed, with workers split on whether $A$ causes $B$ or $B$ causes $A$.
Such a split is a useful signal that more information may be needed to better understand that particular term pair and perhaps other causes and effects related to those terms.

\begin{figure}
\centering
{\includegraphics[width=0.8\textwidth,trim=0 10 0 5,clip=true]{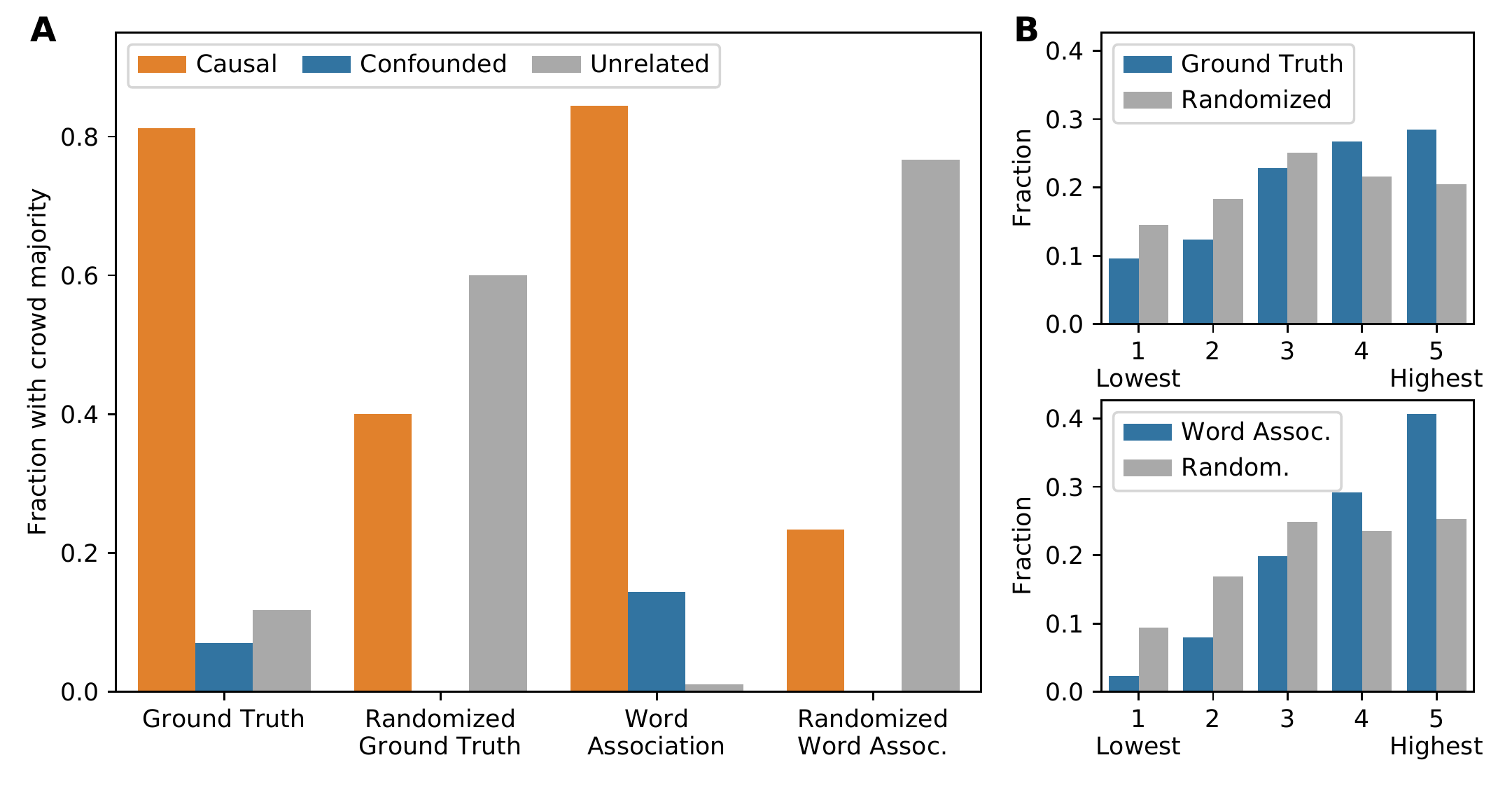}}
\caption{Results of Experiment 1.
\lett{A} The proportion of majority responses for the four datasets.
\lett{B}
Self-reported confidences given by workers for their answers for the four seed datasets.
Confidence was indicated by a 1--5 Likert scale.
\label{fig:exp1results}
}
\end{figure}

Figure~\ref{fig:exp1results} compares the majority crowd classification for terms across the test datasets.
In both non-randomized datasets, the most common classification is `causal', whereas in the two randomized variants, the most common classification is `unrelated'.
Comparing the original and randomized ground truth dataset (Fig.~\ref{fig:exp1results}A), there is a clear difference in the proportion of cause-effect pairs labeled as `causal' versus `unrelated': 80\% of the ground truth pairs have a crowd-majority label of `causal' compared with 40\% for the randomized ground truth data.
At the same time, approximately 10\% of ground truth pairs are labeled as `unrelated' by the crowd majority, compared with 60\% for the randomized ground truth. 
A similar classification difference holds between the word association and randomized word association data (also Fig.~\ref{fig:exp1results}A).

However, workers appear to be over-reporting the incidence of causal relationships: 40\% causal relationships for the randomized ground truth data and 80\% causal relationships for the word association data both seem quite high. 
Workers may be biased in favor of explaining causal relationships where none exist~\cite{taylor1975point,bohner1988triggers}.
At the same time, however, other factors may be at play:
(i) the domain of terms within the ground truth dataset is relatively narrow, making it more likely for random pairs to be related;
(ii) the word association terms are generated by individuals who could very well be using cause-effect reasoning when they ideate an associated term. 
Likely both biased over-reporting and true causal relationships are leading to the causal attribution rates for the shuffled data.

Workers were asked to report their self-confidence in their responses on a 1--5 Likert scale (1-lowest confidence). 
Figure~\ref{fig:exp1results}B reports the distributions of their scores for the first four datasets.
Workers were more confident overall for the word association data than the ground truth data, plausible as the latter were taken from more technical domains. 
Workers responding to randomized variants of both datasets typically had lower confidence than workers responding to pairs taken from the original datasets

\section{Efficient exploration at scale}
\label{sec:algorithms}

A drawback of the single-link learning scheme studied in Experiment 1 is lack of scalability. 
To build a causal network requires asking potentially many workers to support or falsify each link within a single task, which may become cost prohibitive.
To address this, here we introduce a multi-stage algorithm that efficiently explores a network such as a causal attribution network. 
The algorithm proceeds by first asking workers to develop a \textbf{pathway} or chain of causes and effects rooted by an initial seed term.
This pathway is then taken by workers and modified into new pathways by adding new terms, removing old terms, and so forth. 
This process iterates, repeatedly developing new pathways from old ones.
Lastly, the final network is taken from the union of all worker-derived pathways.

We describe the algorithm in the context of causal attributions, but this scheme can be applied with little or not modifications to any network where workers possess enough knowledge and context to offer helpful exploration.
The full algorithm is summarized in Alg.~\ref{alg:algorithm}.
We implement this algorithm in our second experiment (Sec.~\ref{sec:exp2})
and explore how well it samples a known network computationally in Sec.~\ref{sec:model}.

\subsection{Cause Proposal (CP)}

The algorithm begins from a single seed cause. 
A researcher can choose this cause as a starting point based on her needs or research interests, or she can ask the crowd to propose a cause that is of interest to them.
For the latter case, used in this study, the proposal phase can be broken down into two steps: first, workers propose a collection of potential causes, and then workers can rank that list of causes according to some desired criteria.
This ranking can be developed by showing the workers a list and asking for their top choice, asking the workers to sort the list themselves, or by performing pairwise comparisons and then using a ranking algorithm~\cite{dwork2001rank} such as Rank Centrality~\cite{rankcentrality2017} to select the top cause. The latter is most useful if many causes are proposed.
The top ranked cause (or causes) can then be used as the seed for the next phase.

\subsection{Greedy Pathway Expansion (GPE)}

\begin{figure}
\centering
{\includegraphics[width=0.36\textwidth]{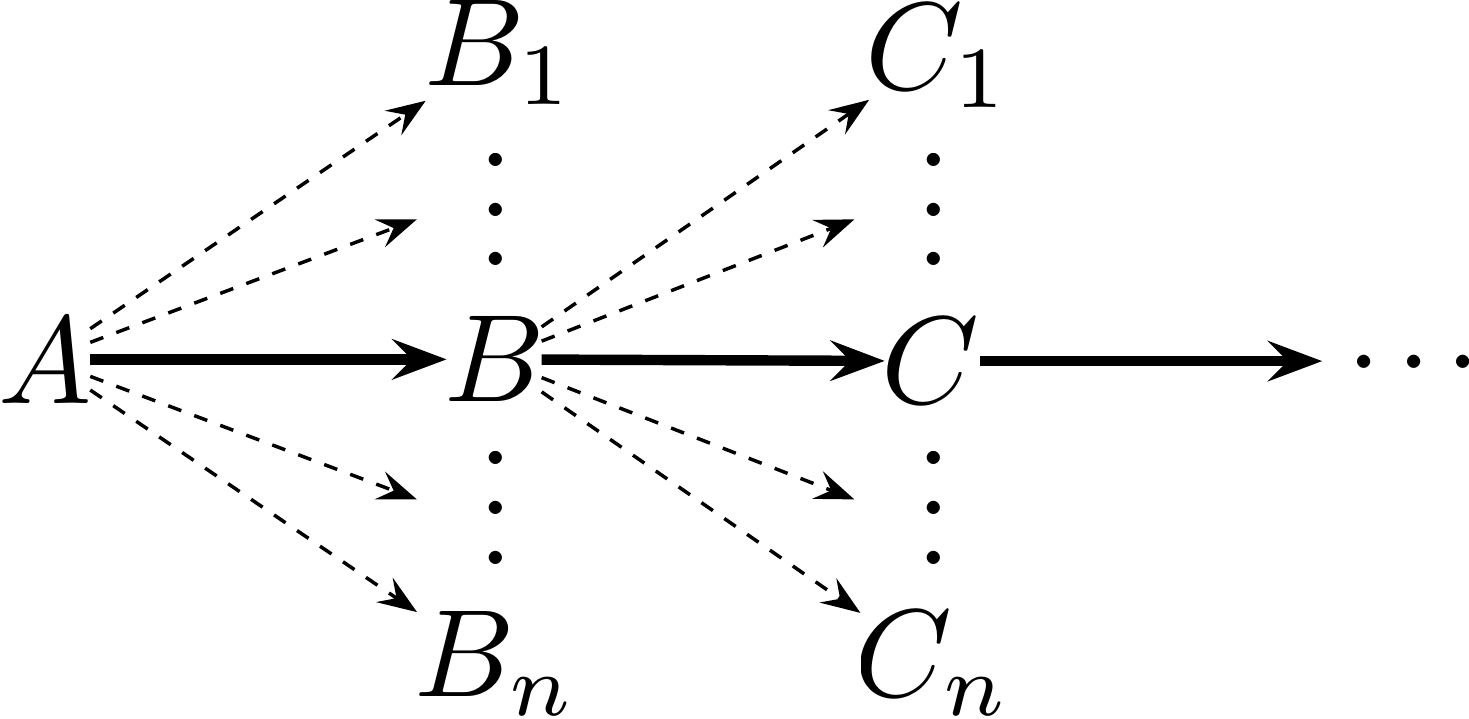}}
\caption{Greedy Pathway Expansion (GPE) to learn the causal chain $A \to B \to C \to \cdots$.
Starting from a seed cause $A$, $n$ candidate effects $B_1, \ldots, B_n$ are proposed and validated by the crowd. The highest quality $B_i$, meaning the effect $B_i$ for which the crowd most agrees on $A \to B$, is then chosen as the first link in the pathway. This $B$ is then used as the start for the next iteration, which continues until $N$ links along the pathway have been determined.
\label{fig:GPE_diagram}
}
\end{figure}

This phase begins from one or more previously developed seed causes.
Workers are asked to propose effects for a given cause (\emph{``What do you think is caused by `poverty'?''}).
The crowdsourcer develops $n$ potential effects $\{B_i\}_{i=1}^{n}$ for cause $A$.
Then, workers are asked to rank these effects from most plausible to least plausible and the top-ranked $B_i \equiv B$ is chosen.
That cause-effect term pair $(A,B)$ forms the first link in a chain of causes and effects.
This process then repeats starting from $B$ to learn the next link, and this continues until a full pathway $P=\left(A, B, C, \ldots \right)$ of length $N$ is developed.
An illustration of this greedy pathway expansion is shown in Fig.~\ref{fig:GPE_diagram}.

These propose-and-rank steps are very similar to those used in the previous proposal phase, and workers can easily be shared between these tasks, although one may wish to use different sets of workers for different phases if desired, for example, if there is a concern about possible bias in their responses.

It is also worth noting that workers asked to rank effects will see each link of the pathway independently, meaning they are asked to evaluate potential link $(A, B_i)$, link $(B, C_i)$, etc. 
They do not see, for example, the longer chain of causes and effects $(A,B, C_i)$, and seeing the context of $A$ may potentially change their selection of the best candidate effect $C_i$ for a given cause $B$.
The next phase of our algorithm addresses this, by allowing workers to see and modify longer pathways of causes and effects.

\subsection{Iterative Pathway Refinement (IPR)}

The goal of this phase of the crowdsourcing is to develop a larger set of pathways starting from the pathway(s) developed during the previous GPE phase.
Here workers are shown a pathway using a web form that allows them to quickly edit and update the pathway.
Updates can include insertions, deletions or rearrangements of the terms within the pathway, and workers can accomplish all of these tasks with a simple drag-and-drop interface in order to be as efficient as possible.

Each worker participating in this phase is shown a chosen pathway $P_i$, either the original GPE pathway or a subsequently developed pathway introduced by an earlier IPR worker. 
The worker uses the web form to modify $P_i$ into a new pathway $P_{i+1}$. 
This new pathway is inserted back into the pathway set and the algorithm repeats for the next worker.

Because workers develop entire cause-effect pathways per microtask, this task structure provides considerably more information about the network per task than could be attained by the single-link  (SL) learning experiment (Experiment 1). 
Of course, IPR is a more complex task than SL, so it is important to assess the \emph{speed} of workers when investigating the efficiency of IPR, which we do in Experiment 2 (Sec.~\ref{sec:exp2}).
By taking the union of all the pathways, a single causal attribution network can be estimated.
Specifics details on network extraction are given in Algorithm~\ref{alg:algorithm}.

\begin{algorithm}[t]
\begin{enumerate}\itemsep=0.667em
\item Cause Proposal (CP):
    \begin{enumerate}\itemsep=\myitemsep
    	\item Workers propose $n$ root causes $A_i$.
    	\item Workers rank these $A_i$'s by interest, select top ranked $A \in \{A_i\}_{i=1}^{n}$ as the seed cause.
    \end{enumerate}

\item Greedy Pathway Expansion (GPE) (Fig.~\ref{fig:GPE_diagram}):
    \begin{enumerate}\itemsep=\myitemsep
    	\item Workers propose candidate effects $B_i$ for seed cause $A$, leading to potential causal links $A \to B_i$\label{item:item1}.
    	\item Workers vote on these $A \to B_i$. Choose the top ranked $B \in \{B_i\}_{i=1}^{n}$, leading to causal link $A \to B$.
    	\item Repeat from \ref{item:item1} with $B$ as the new root cause until a pathway $P_0 = \left(A, B, \ldots\right)$ of length $N$ is achieved.
    \end{enumerate}

\item Iterative Pathway Refinement (IPR):\label{item:anotheritem}
    \begin{enumerate}\itemsep=\myitemsep
    	\item Initialize a set of pathways $\mathcal{P}$ seeded by the initial GPE pathway, i.e.\ $\mathcal{P} = \{P_0\}$.
    	\item Workers iteratively refine pathways. A new worker is shown a selected pathway $P_i \in \mathcal{P}$, and uses a drag-and-drop interface (Fig.~\ref{fig:IPRformscreenshot}) to add, remove, and reorder terms in $P_i$ to  create $P_{i+1}$. The new pathway is added to the set: $\mathcal{P} \leftarrow \mathcal{P} \cup \{P_{i+1}\}$\label{item:item2}.
		\item Repeat from \ref{item:item2} until a desired number of pathways are developed.
    \end{enumerate}
   
\item  Network Extraction:
	\begin{enumerate}\itemsep=\myitemsep
	\item Define a network $G = (V,E)$ by taking the union of all pathways developed in \ref{item:anotheritem}. Specifically, 
	$V = \{T \mid T \in P_i, \mbox{~where~} P_i \in \mathcal{P}\}$ and
	$E = \{ (T_j, T_{j+1}) \mid T_j, T_{j+1} \in V, \exists P_i \in \mathcal{P} \mbox{~s.t.~} (T_j,T_{j+1}) \mbox{~is a substring of~} P_i \}$.
	\end{enumerate}
\end{enumerate}
\caption{Multi-phase crowdsourcing algorithm for (causal) network exploration. All phases of the algorithm can run in parallel: the crowd can develop multiple seed causes, work simultaneously on multiple GPE pathways, and so forth.
\label{alg:algorithm}
}
\end{algorithm}

\subsection*{Extensions and modifications}

The multi-phase algorithm we propose here is modular and extensible in several ways, and a crowdsourcer is free to adapt the different components to meet her needs.
For example, a crowdsourcer may skip the proposal of root terms if she already has a seed set to use. 
Likewise, one can bypass the Greedy Pathway Expansion phase and go directly to Iterative Pathway Refinement, either by using a pathway of one term as the original pathway seed set, or by starting from one or more predetermined pathways if they are available.
The algorithm is easily parallelized, allowing the same crowd to contribute simultaneously to different network explorations, for example starting from different seeds.
Lastly, while the algorithm is described and applied in this work to the case of directed causal attribution networks, it is by no means limited to such cases, and can serve as an effective crowd exploration method for any networks where the crowd is suitable for exploration.
Indeed, Iterative Pathway Refinement is a special case of a more general exploration method---\emph{Iterative Motif Refinement}---where workers construct and modify small subgraphs known as motifs~\cite{milo2002network}, of which directed pathways are one such motif. 
We discuss such generalizations further in Sec.~\ref{sec:discussion}.

\section{Experiment 2 --- causal attribution at scale}
\label{sec:exp2}

To demonstrate our crowdsourcing algorithm (Sec.~\ref{sec:algorithms}), we implemented it on Amazon Mechanical Turk (AMT) as a collection of interrelated Human Intelligence Tasks (HITs).
Our implementation proceeded primarily along the lines of the algorithm delineated in Sec.~\ref{sec:algorithms}, with a few practically-motivated key differences. 
For example, multiple causes were proposed via the Cause Proposal HITs and multiple pathways were constructed simultaneously during the Greedy Pathway Expansion HITs.

\subsection{Materials and methods}

We implemented three HIT web interfaces for workers corresponding to the Cause Proposal (CP), Greedy Pathway Expansion (GPE), and Iterative Pathway Refinement (IPR) phases of the crowdsourcing algorithm detailed in Sec.~\ref{sec:algorithms}.
These are known as ``external'' HITs on AMT as the web forms are hosted on our own server that workers access through an iframe within the AMT website.
At each phase we chose various criteria described below for the number of responses to collect, balancing the quantity of data needed with budgetary limits.
Practitioners applying our algorithm or one similar to it will likely face similar decisions but this will generally depend on their particular circumstances.
For all tasks, workers could not respond multiple times to the same information; they could not vote on the same cause-effect pair more than once, for example.
%
This research procedure has been approved by our IRB (\detnum{}).

The CP task showed workers an example of a cause-effect pathway and asked the worker either to propose or to rank 3--5 causes that they believed ``may have very surprising or unintended or interesting effects'' (see App.~\ref{app:mturkInstructions} for screenshots of task interfaces with instructions).
Care was taken with these instructions: we wanted the workers to have simple instructions that demonstrated causal relationships but did not prime them in terms of quantifying how interesting or important the causes should be.
While we could have given them detailed selection criteria to follow (we plan to study this in future work), here we are mainly interested in determining what crowd workers consider important, as free from our influence as possible.
This goal is in contrast with most crowdsourcing research that emphasizes the importance of detailed, exact instructions.
In this phase, we sought 50 proposed causes, 20 worker rankings per cause, and we selected the top eight ranked causes. These eight root or seed causes were then passed to the GPE HITs.

For the GPE task, we used a HIT very similar to the CP HIT but now asking workers to either (i) propose effects for a given cause, or (ii) vote on effects by either agreeing or disagreeing with a given cause-effect pair. See App.~\ref{app:mturkInstructions} for the GPE instructions.
At each step when growing a GPE pathway from a given cause, we sought 10 unique effects for that cause (part i), and we sought 5 different workers to validate each of the 10 effects (part ii).
When finished, the effect which received the highest proportion of affirmative votes was selected as the next piece of the GPE pathway (any ties were broken at random) and the GPE task restarted with the newly chosen effect as the given cause.
Note that the GPE phase was conducted on all 8 pathways in parallel, and workers could propose and/or vote on any tasks available in any of the chains.
We continued this process until the GPE pathways had an average length of 5 terms.
These pathways then seed the IPR HITs.

\begin{figure}
\centering
{\includegraphics[width=0.6\textwidth]{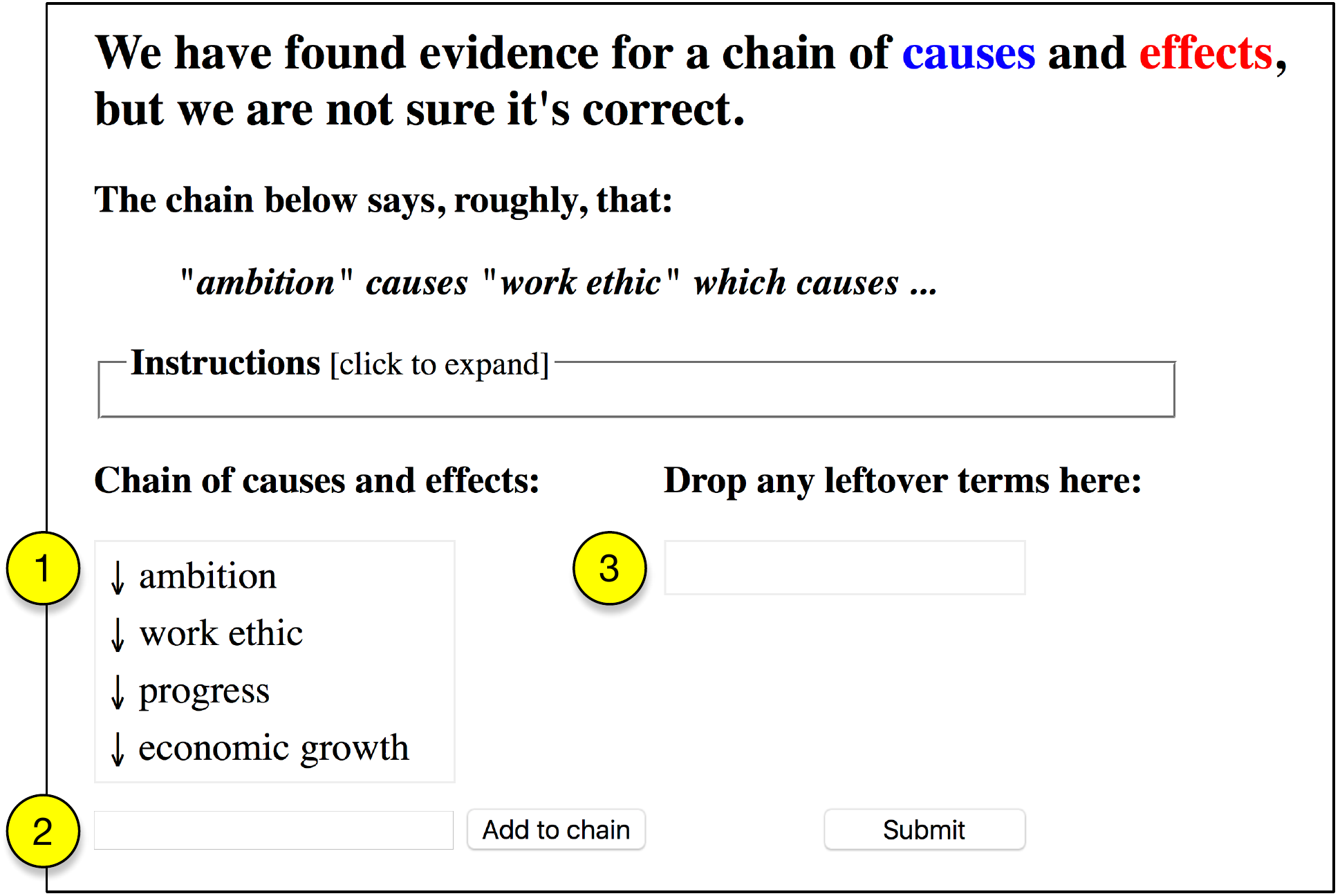}}
\caption{Screenshot of the task interface for workers participating in Experiment 2's Iterative Pathway Refinement task. The pathway of terms listed in the lower left (Callout 1) is reorderable using drag-and-drop operations. Insertions and deletions can be made using the form as well (Callouts 2 and 3, respectively). The instructions box is collapsed in this view but is shown expanded before the worker chooses to accept the HIT (see Appendix~\ref{app:mturkInstructions}).
\label{fig:IPRformscreenshot}
}
\end{figure}

The IPR task is the most important phase of our algorithm, and subsequently had the most detailed, interactive web interface.
Figure~\ref{fig:IPRformscreenshot} shows a screenshot of the IPR interface.
Unlike tasks in the previous phases, there is no switching between propose and vote subtasks.
Workers were simply shown a causal pathway selected at random from the set of available pathways and asked to modify the pathway. 
The pathway was presented as a vertical list that workers could rearrange using drag-and-drop operations. Workers could also insert terms into the pathway and delete terms from the pathway using the same task interface.
Upon receiving the modified pathway from the worker, we inserted it back into the pool of pathways alongside the original pathway.
To prevent workers from modifying old pathways too often, pathways are flagged as `unavailable' if more than 5 modified pathways were derived from it by workers.
We sought 1500 IPR responses from workers.

\subsection{Results}

The three phases of the crowdsourcing algorithm (Cause Propose, Greedy Pathway Expansion, Iterative Pathway Refinement) were run sequentially on Amazon Mechanical Turk.
We did not filter or manually remove any responses across the tasks. However, there is often a lag between giving a worker a task and receiving a response, so we often receive slightly more responses than necessary as the ``stragglers trickle in''. 
We consider this acceptable as it has a minor effect on our budget, but in principle a crowdsourcer could more carefully reserve tasks deployed to workers to prevent this.
Altogether, 27 workers submitted 48 unique (50 total) root causes.
We selected the top 8 ranked causes to initialize 8 GPE pathways. 
These pathways were then expanded by workers until they had an average length of 5 terms; we received 41 distinct terms across the 8 pathways.
Lastly, IPR workers provided 1567 pathways, including the 8 original GPE pathways.

The net result is a collection of 1567 pathways representing 394 unique terms and 1329 distinct cause-effect term pairs. 
We summarize the numbers of responses, numbers of workers, and rewards per HIT for each phase in Table~\ref{tab:exp2taskSummary}.

\begin{table}[t]
\caption{Summary of crowdsourcing tasks for Experiment 2. Rewards are in USD.}
\begin{center}\small
\begin{tabular}{llll}
Task & \# responses & \# workers & reward per HIT\\
\hline
Cause Proposal:              &           &       & \\
--- Proposal                 &   17      & 17    & \$0.17     \\
--- Rank                     &   926     & 98    & \$0.17      \\
Greedy Pathway Expansion:    &           &       &  \\
--- Proposal                 &   591     & 84    & \$0.07     \\
--- Vote                     &   2136    & 122   & \$0.07     \\
Iterative Pathway Refinement &   1567    & 96    & \$0.17
\end{tabular}
\end{center}
\label{tab:exp2taskSummary}
\end{table}%

Likewise, Table~\ref{tab:GPEresultsExp2} summarizes the eight initial seeds and subsequent pathways built by crowd workers during
the first two phases of the experiment. The first term of each pathway is a seed developed during the first phase, while the pathway itself was developed during the second or GPE phase.
All terms here and throughout the causal network were introduced by workers.
Interestingly, there is a roughly even split between positive sentiment (``ambition'', ``education'') and negative sentiment (``inequality'', ``fear'') seed terms.

\begin{table}[t]
\caption{Initial pathways developed using Greedy Pathway Expansion in parallel during Experiment 2. The seed terms developed during the Cause Proposal task are the first terms of these pathways. All terms were introduced by workers. Pathways are listed in arbitrary order.}
\begin{center}\small
\begin{tabular}{rp{5in}}
\hangindent=1em
1) & inequality $\to$ resentment $\to$ contempt $\to$ anger $\to$ mistakes $\to$ accidents $\to$ personal injury $\to$ pain $\to$ suffering\\
2)  & wealth $\to$ stability $\to$ comfort $\to$ security\\
3) & poverty $\to$ hunger $\to$ starvation $\to$ death suffering misery $\to$ sadness\\
4) & fear $\to$ sweating $\to$ dampness $\to$ mold $\to$ illness\\
5) & curiosity $\to$ ideas $\to$ inventions $\to$ new products\\
6) & ignorance $\to$ stupidity $\to$ bad mistakes $\to$ bad results\\
7) & ambition $\to$ work ethic $\to$ economic growth $\to$ progress $\to$ growth\\
8) & education $\to$ knowledge $\to$ curiosity $\to$ discovery $\to$ invention growth change
\end{tabular}
\end{center}
\label{tab:GPEresultsExp2}
\end{table}%

The GPE pathways shown in Table~\ref{tab:GPEresultsExp2} were used to seed the pathway set developed during the Iterative Pathway Refinement (IPR) phase. 
Workers in this phase developed a total of 1567 distinct causal attribution pathways.
Only 5 pathways contained a duplicate term and only one was a self-loop (``economic growth'' caused ``economic growth'').

\subsubsection{\textbf{Patterns of pathway refinement}}

\begin{figure}
\centering
{\includegraphics[width=0.75\textwidth,trim=0 10 0 6,clip=true]{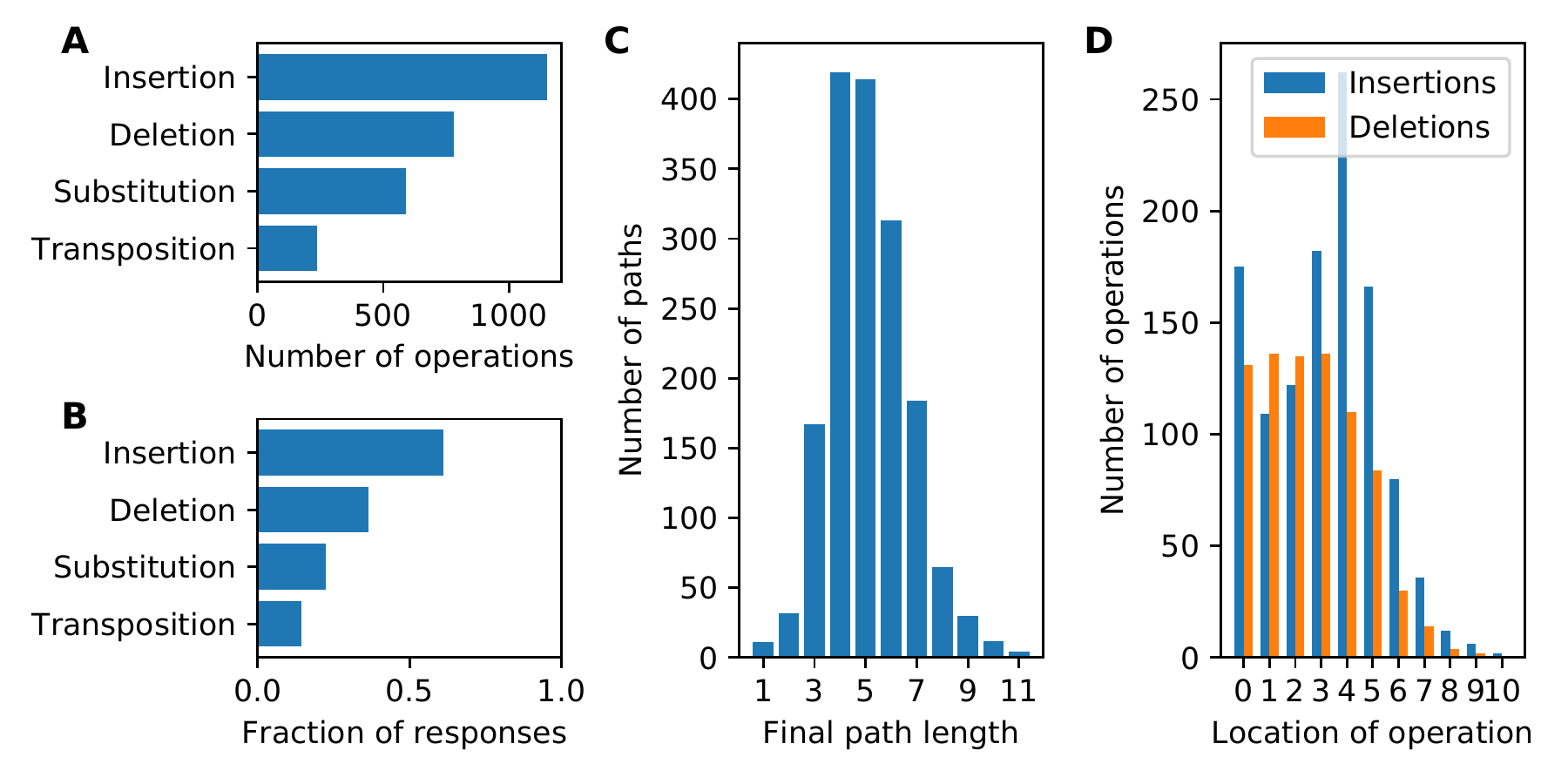}}
\caption{Iterative Pathway Refinement (IPR) edit operations and cause-effect pathway lengths.
\lett{A} The numbers of each type of IPR edit operation.
\lett{B} The fraction of responses containing at least one edit operation.
\lett{C} The distribution of pathway lengths.
\lett{D} The distributions of locations of the two most common edit operations.
\label{fig:microPathStats}
}
\end{figure}

We explore the patterns of pathways and how workers edit pathways in Fig.~\ref{fig:microPathStats}.
Using the IPR interface, workers given a path can make any of four \emph{\textbf{edit operations}}: insertion of a new term, deletion of an old term, substitution of an existing term with a new term (counted as a separate operation although it is a deletion and subsequent insertion at the same position), or the transposition of two existing terms.
The first two edit operations are the most common choices workers make (Fig.~\ref{fig:microPathStats}A, B).
At least one insertion occurred in over 50\% of responses while at least one deletion occurred in over a third of responses (Fig.~\ref{fig:microPathStats}B).
The mean length of pathways submitted by workers is 5.17 terms; the distribution of pathway lengths is shown in Fig.~\ref{fig:microPathStats}C.

Tracking the locations of the two most common edit operations performed by workers, insertions and deletions, both operations are more likely to occur towards the beginning of a given pathway than expected from the overall distribution of pathway length (Fig.~\ref{fig:microPathStats}D). 
Further, insertions occur more often than deletions at the end of pathways.

A bias in favor of edits located near the beginning of a pathway could be the result of cognitive mechanisms by which individuals process chains of causes and effects. 
But it could instead simply be evidence that workers are \emph{\textbf{satisficing}}, finding means to acceptably complete the IPR task as quickly as possible. 
If the latter, this could indicate that workers are preferentially ignoring the later terms in a pathway.

\begin{figure}
\centering

{\includegraphics[width=0.8\textwidth,trim=0 10 0 7,clip=true]{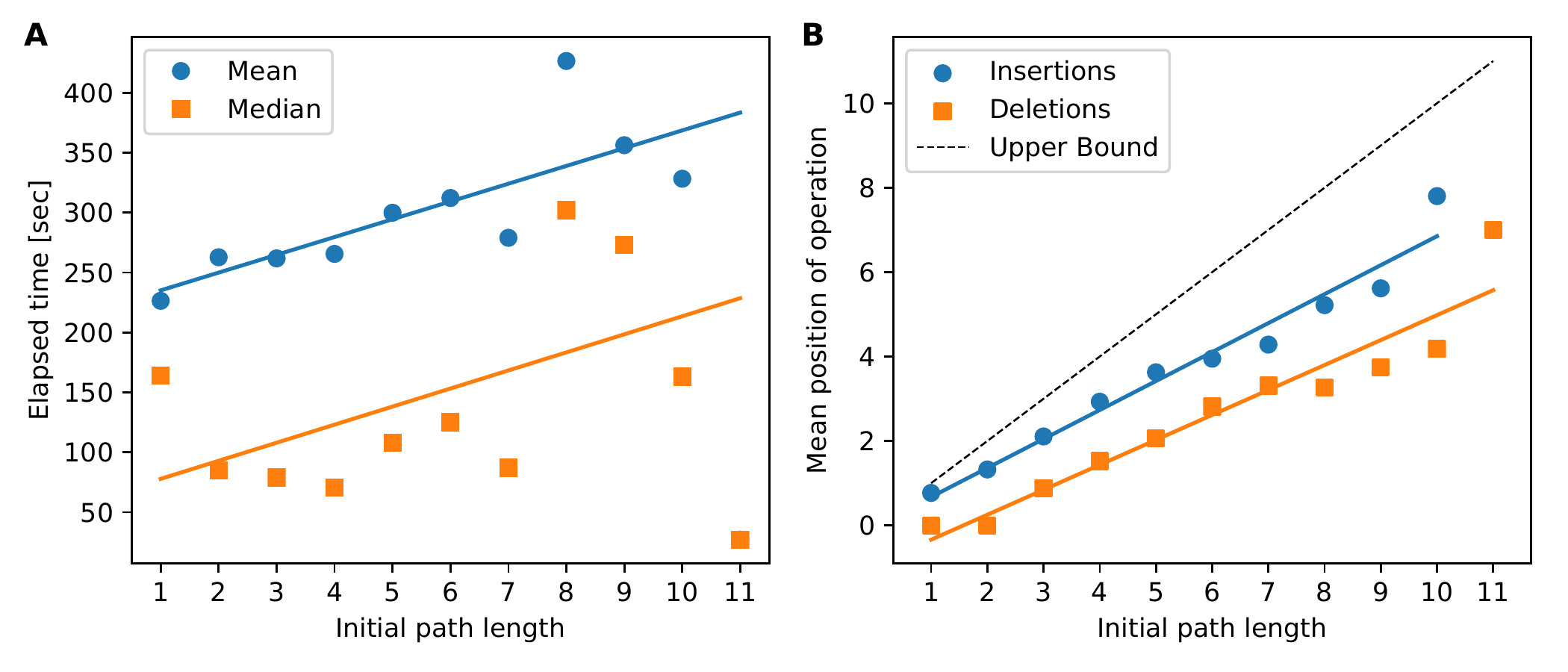}}
\caption{Most workers do not ``satisfice'' the IPR task by focusing only at the start of a pathway, even when shown longer pathways.
\lett{A} The estimated elapsed time (in seconds) for IPR workers to submit their response as a function of
the length of the path they are presented.
\lett{B} The average location of edit operations grows as the initial path length increases.
Best-fit lines provide guides for the eye.
\label{fig:pathwaySatisficing}
}
\end{figure}

To investigate satisficing, in Fig.~\ref{fig:pathwaySatisficing} we study the elapsed time workers spend on a given IPR task and the average positions of insertions and deletions within the pathway, both as functions of the initial length of the pathway shown for that task.
If workers are satisficing we expect to observe an approximately flat trend in elapsed time versus initial path length, at least for longer pathways.
Likewise, if satisficing, the average position of insertions and deletions would also tend towards a constant with pathway length.
Neither occurs: the elapsed time grows roughly linearly with initial path length (Fig.~\ref{fig:pathwaySatisficing}A) as does the mean location of edits (Fig.~\ref{fig:pathwaySatisficing}B). 
Further, if some workers were satisficing by only studying the tail of the pathway instead of the head, we would observe a non-monotonic trend in Fig.~\ref{fig:pathwaySatisficing}, which does not occur.
While satisficing is likely still occurring for some workers, these trends provide evidence that many if not most workers are taking the time to process and understand the full length of the pathway.

\subsubsection{\textbf{The causal attribution network}}
\label{subsubsec:thecausalnetwork}

\begin{figure}
\centering
{\includegraphics[width=\textwidth]{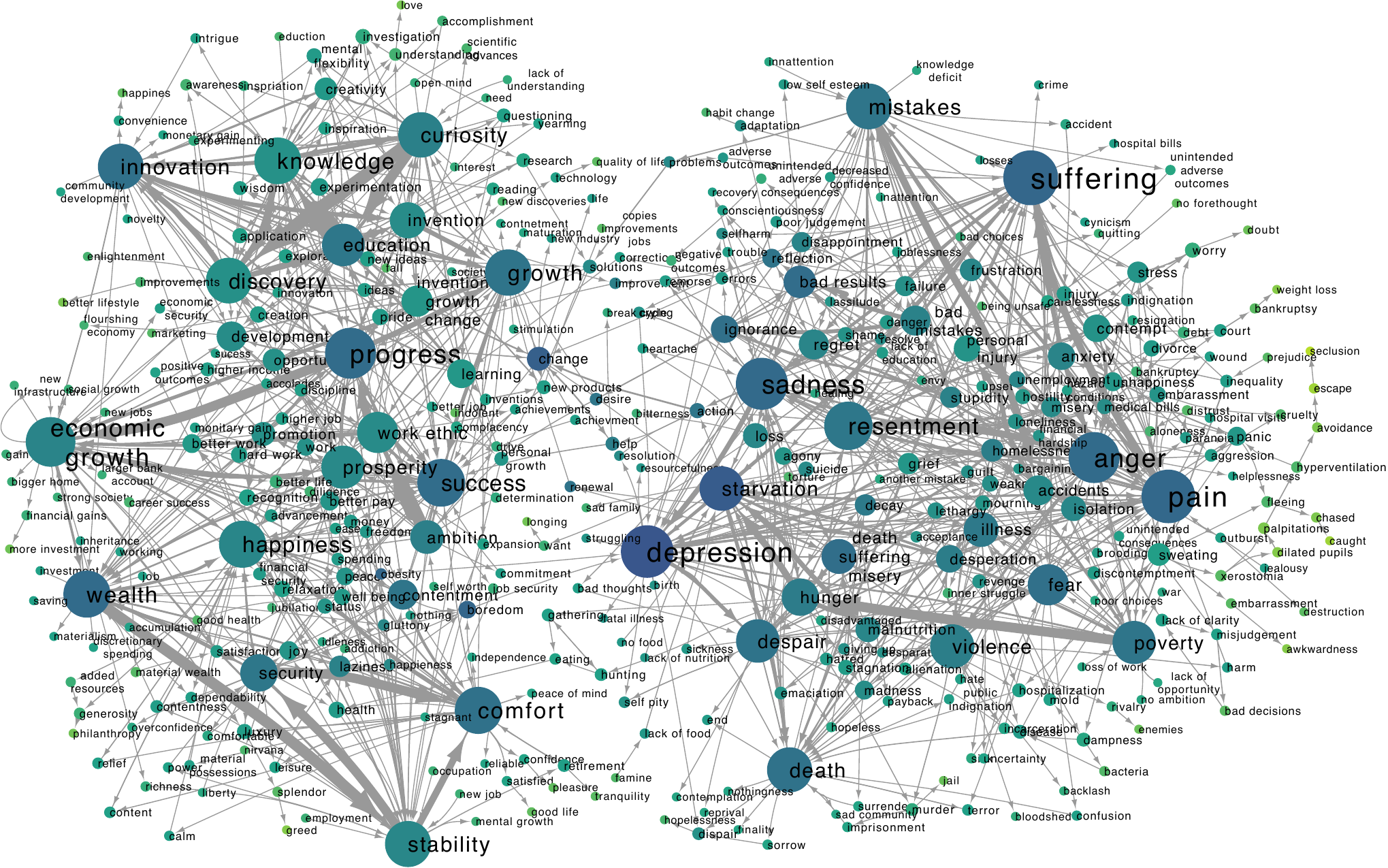}}
\caption{A network visualization of the directed causal attribution network constructed by workers. 
Node size is proportional to degree (indegree $+$ outdegree), node color proportional to closeness centrality, edge width proportional to edge weight.
\label{fig:exp2network}
}
\end{figure}

Taking the union of all IPR pathways in the Network Extraction phase (Sec.~\ref{sec:algorithms}) yields the final causal attribution network developed by this experiment. 
This is a weighted, directed network where each edge ($i,j$) indicates that $i$ ``causes'' $j$, and the edge weight $w_{ij}$ associated with this edge denotes the number of IPR pathways containing the directed link ($i,j$).
We visualize this network in Fig.~\ref{fig:exp2network}.

We remark upon a number of interesting features of this causal network. 
Particularly intriguing is that the network is roughly clustered into two groups, one focused on positive terms (``progress'', ``growth'', ``success'', etc.), and another around negative terms (``suffering'', ``resentment'', ``depression'').
This clustering, which comes from the initial GPE pathways (Table ~\ref{tab:GPEresultsExp2}) likely reflects known emotional and valence effects underlying causal attribution~\cite{bohner1988triggers}. 
We see that workers are more likely to ideate causal relationships that are at the emotional extremes, either very positive or very negative.

Within the clusters, multiple causal links appear strongly supported by the crowd, for example, the links between stability, security, and comfort in the lower left of the network or the link from poverty to hunger (and not vice versa). 
There are also a number of cause-effect pairs where the crowd is split on the directionality: for example, ``discovery causes knowledge'' occurs in 57 pathways while ``knowledge causes discovery'' occurs in 64 pathways.

\begin{figure}
\centering
{\includegraphics[width=0.8\textwidth,trim=0 10 0 0,clip=true]{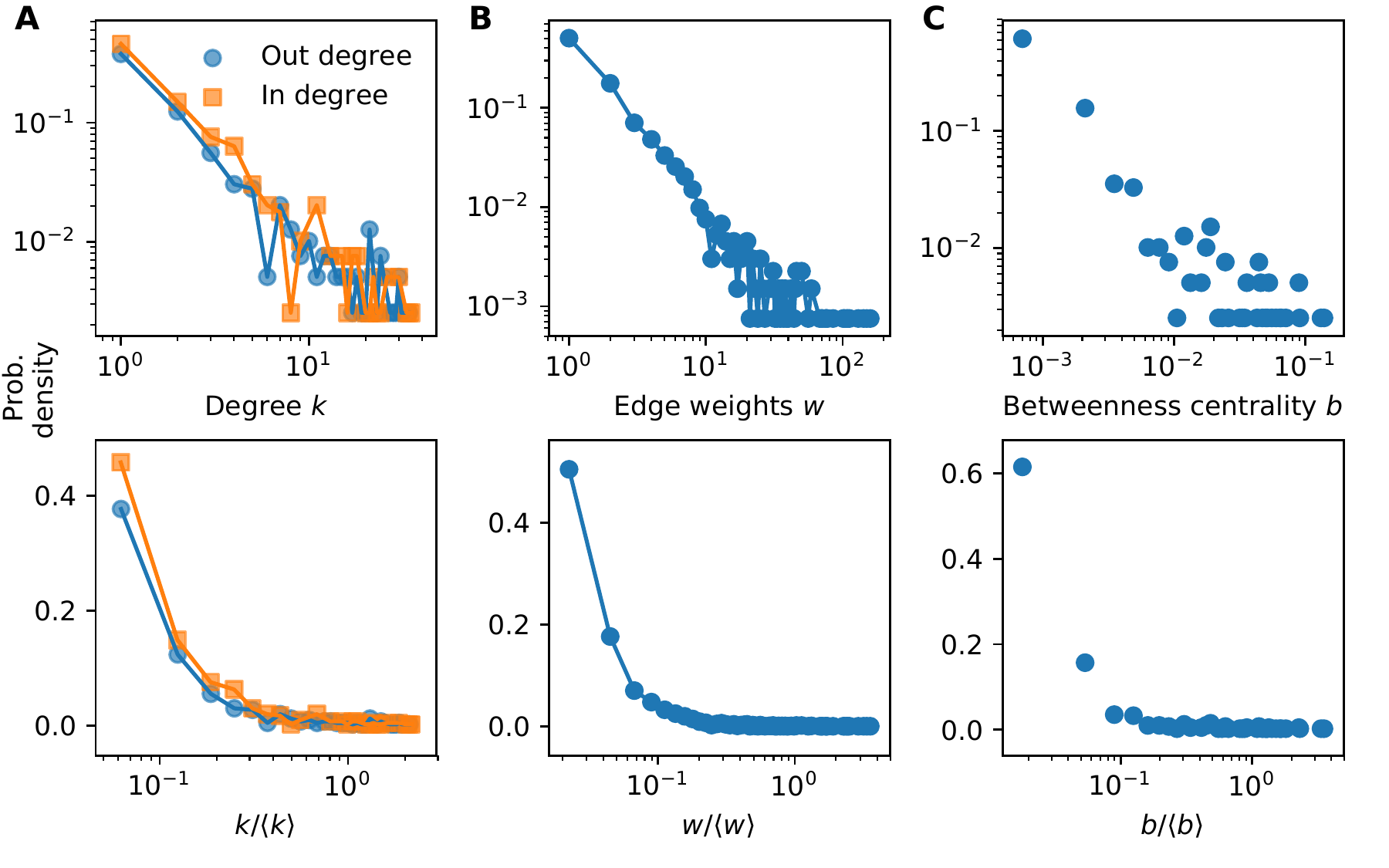}}
\caption{Properties of the crowdsourced causal network shown in Fig.~\ref{fig:exp2network}.
\lett{A} Degree distributions,
\lett{B} edge weight distribution, 
\lett{C} Betweenness centrality distribution.
\label{fig:graphDistributions}
}
\end{figure}

Turning our attention from the qualitative view of the network (Fig.~\ref{fig:exp2network}), we next investigate quantitative properties of the network. Figure~\ref{fig:graphDistributions} shows the distributions of indegree, outdegree, edge weight (number of pathways containing a given cause-effect pair) and betweenness centrality (how often a term is in a shortest path in the network). In all cases, these statistics follow heavy-tailed, non-Gaussian distributions.
Interestingly, after rescaling each quantity by its mean value (Fig \ref{fig:graphDistributions}, bottom row), the different quantities follow similar distributions.

\begin{figure}
\centering
{\includegraphics[width=0.75\textwidth,trim=0 5 0 0,clip=true]{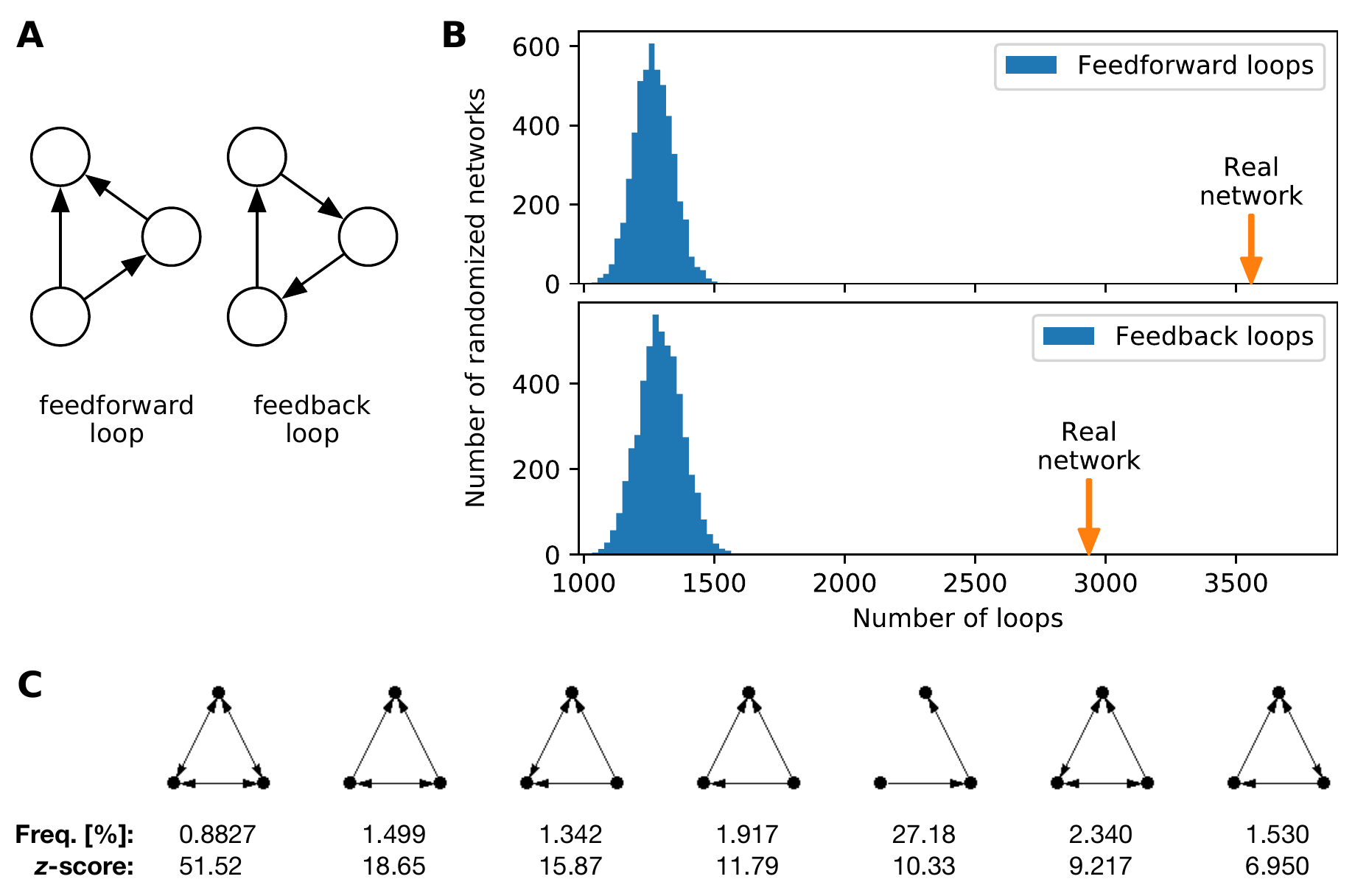}}
\caption{The crowdsourced causal attribution network has significant feedforward loop, feedback loop, and other motif structure.
\label{fig:feedforward}
}
\end{figure}

Lastly, to evaluate the significance of the causal network structure, at least along one dimension, we consider the prevalence of \emph{\textbf{feedforward loops}} and \emph{\textbf{feedback loops}} in the network (Fig.~\ref{fig:feedforward}A). 
Feedforward loops are an important structural motif in directed networks~\cite{milo2002network}
and feedback loops are a key component of causal homeostasis~\cite{keil2006explanation}.
We used the statistical procedure of Milo \emph{et al.}~\cite{milo2002network} to evaluate the significance of the three-node feedforward loop and feedback loop motifs. 
Specifically, we enumerated all feedforward and feedback loops in the crowdsourced network and compared that
number in Fig.~\ref{fig:feedforward}B with the same quantity computed from 5000 null networks generated using an in- and outdegree-preserving randomization procedure~\cite{milo2002network}.
The observed network has significantly more feedforward loops and feedback loops than expected according to Milo \emph{et al.}'s null model ($z \approx 30.99$, $p < 10^{-200}$ for feedforward loops; $z \approx 19.30$, $p < 10^{-80}$ for feedback loops).
Expanding on our analysis of feedforward and feedback loops, Fig.~\ref{fig:feedforward}C shows all the statistically significant three-node \emph{motifs}, including their occurrence frequency and $z$-scores relative to the null model. Seven of the 16 possible three-node motifs were significant. 
Motifs in Fig.~\ref{fig:feedforward}C were found using FANMOD with default parameters~\cite{wernicke2006fanmod}.
Understanding these structural motifs allows us to better characterize properties of the crowdsourced causal attribution network.

\subsubsection{\textbf{Speed of workers}}
\label{subsec:speedIPRworkers}

\begin{figure}
\centering
{\includegraphics[width=0.5\textwidth,trim=0 15 0 10,clip=true]{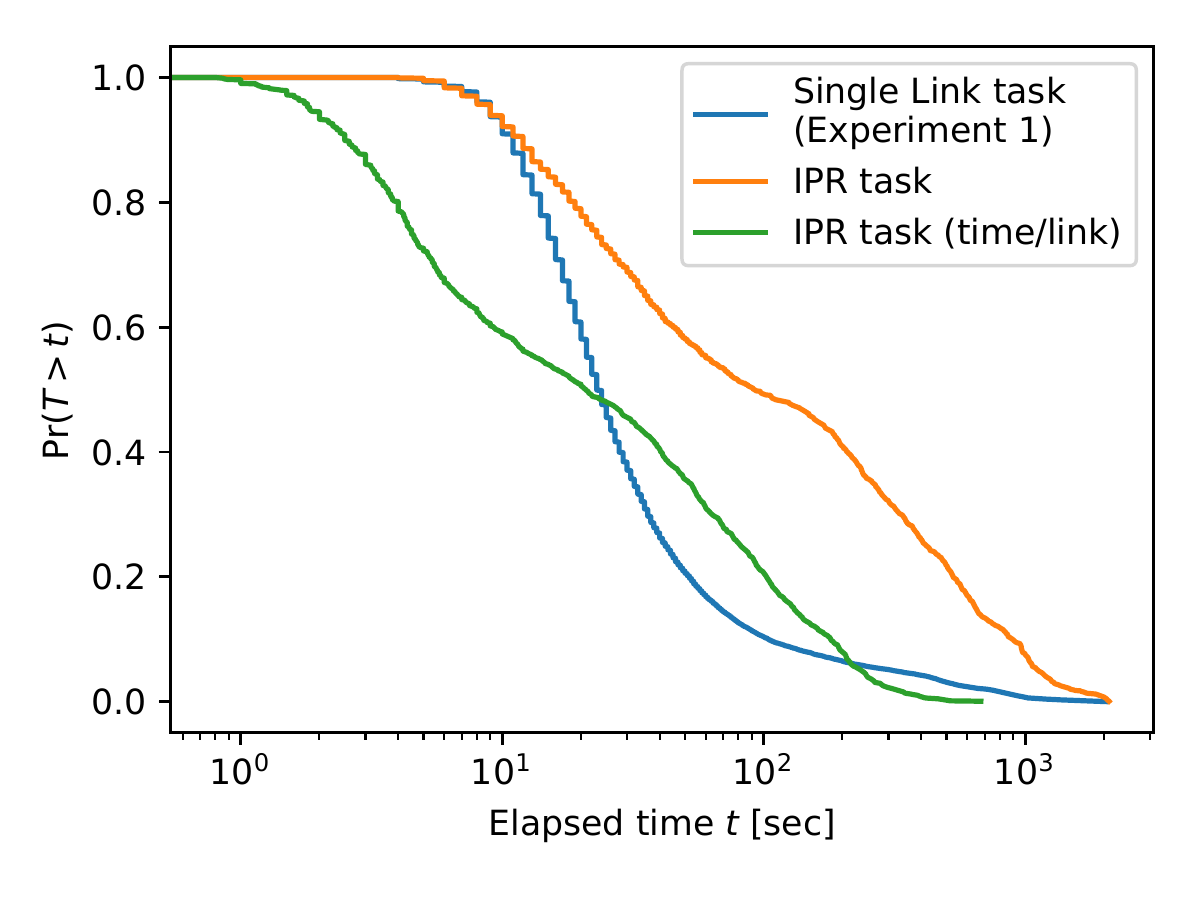}}
\caption{Cumulative distributions of the elapsed time (as reported by AMT) for workers to complete tasks.
\label{fig:elapsedTimes}
}
\end{figure}

The IPR task will explore more of a network with fewer worker responses than the ``single link'' task used in Experiment 1 (Sec.~\ref{sec:exp1}).
But this is expected, as each IPR task is more complex and provides more information.
But since the IPR task is so complex, it is likely that workers require more time to complete IPR tasks than  single link tasks.
Indeed, examining in Fig.~\ref{fig:elapsedTimes} the distributions of \emph{\textbf{elapsed times}}, the delay between when a worker receives and submits a task shows exactly that: IPR workers require more time than single link workers to complete their task.

However, while slower elapsed times for the IPR task seems to imply that the crowdsourcing algorithm is not efficient in terms of time-to-explore, the direct comparison between these elapsed times is not necessarily appropriate.
As workers provide information on multiple links within a single IPR task, it is better to compare the elapsed time \emph{\textbf{per link}} for IPR workers with the elapsed time of single link workers.
Even those links in the pathway that remain unmodified by the worker still receive an additional ``vote'' when taking the union of the pathway set, so normalizing the elapsed time relative to the total length of the IPR pathway accounts for the total information provided by the worker.
Figure \ref{fig:elapsedTimes} shows that IPR workers provide information more quickly per link than single link workers.
IPR workers had a mean elapsed time per link of $59.00$ seconds, compared with $68.03$ seconds on average for single link workers.
This difference in distributions was significant (Mann-Whitney U test, $p < 10^{-15}$).
Thus, the rate of information gain for IPR workers is significantly faster than single link workers.

\subsection{Experiment 2 Discussion}

Taken together, the crowd was able to generate a relatively large and meaningful causal attribution network using our multi-phase algorithm.
Of course, the causal network as it stands after this experiment is not the final form of these data, and more work should be taken to filter the network. 
Specifically, terms that are overly vague (``unintended adverse outcomes'') should be improved, the network likely contains synonymous terms that should be aggregated, and there remain a large number of terms on the periphery of the network that should be further explored. 
That said, the data collected from workers provide useful cause-and-effect relationships, possesses statistically significant network structure, and the edit patterns of workers can help researchers further understand cognitive aspects of causal attribution.

The IPR task forming the core of the crowdsourcing algorithm enabled workers to be significantly faster at providing network information than workers asked to validate a single link per task, with IPR workers completing the task approximately 9 seconds faster per link on average.
Such a time difference when compounded over a large network exploration can have a sizable impact on the overall completion time of the crowdsourcing.
Further improvements on the interface of the IPR task, such as allowing workers to edit pathway terms in place instead of using drag-and-drop operations to update terms, will likely improve worker speed even more.

\section[]{Evaluating response quality}
\label{sec:followupsurvey}

Experiment 2 (Sec.~\ref{sec:exp2}) provides evidence that workers can build a graph more efficiently with our algorithm (Sec.~\ref{sec:algorithms}) than they could working one link at a time (as per Experiment 1 in Sec.~\ref{sec:exp1}). 
But does this efficiency come at a cost? Are we trading off quality for quantity?

While Experiment 1 studies how workers attribute cause and effect using a small ground truth dataset, it is challenging to systematically investigate the quality of Experiment 2's network as no ground truth is available.
In light of this and to provide at least some information on quality, we performed a followup crowdsourcing survey asking workers to examine results generated during Experiment 2.
We presented new crowd workers with potential cause-effect pairs taken from the causal attribution network (using questions of the form ``\emph{Do you think that $A$ causes $B$?}'') and asked them if they agree or disagree (or were uncertain) with these attributions.
The goal of this survey is for workers to signal some degree of relative quality (or at least consistency) by assessing previously generated causal attributions.
This survey task is identical in form to the `vote' phase of Greedy Pathway Expansion (Sec.~\ref{sec:algorithms}).
To provide an independent assessment,
workers who participated in Experiment 2 were excluded from this followup task.

For this new task we extracted two groups of $n = 50$ cause-effect pairs ($n = 100$ total) sampled at random from Experiment 2's graph:

\begin{enumerate}\itemsep=0pt

\item In the first group, each cause-effect pair $(i, j)$ was a directed link within the causal attribution graph. Half ($n = 25$) of the sampled links had low weight ($w_{ij} \leq$ median link weight) while half had high weight ($w_{ij} >$ median link weight).
Since link weight $w$ is the number of worker responses containing the link, 
we assume ``stronger'' links are more likely to be approved.

\item Our second group focuses on disconnected nodes, asking workers to agree/disagree with potential causal attributions not captured in the current network.
Half ($n=25$) of the node pairs were chosen at random from those pairs at distance $d=2$ in the network, meaning they were only two hops apart on the network.
It is plausible that some or even many of these pairs have a causal relationship yet to be introduced by workers since, if the link between them existed, they would form a feedforward loop (Sec.~\ref{subsubsec:thecausalnetwork}).
The second half of the group consists of pairs of nodes at distance $d>2$, capturing more distant nodes.
We expect these node pairs to be less likely to hold a causal relationship, although some may have a relationship.
\end{enumerate}

We collected 984 responses from 114 workers with each of the $n=100$ cause-effect pairs examined by at least 8 workers. Workers were compensated \$0.07 per response and each worker was limited to at most 25 responses to provide room for many different workers to respond.
This research procedure has been approved by our IRB (\detnum{}).

\begin{figure}
\centering
{\includegraphics[width=0.9\textwidth,trim=0 10 0 10,clip=true]{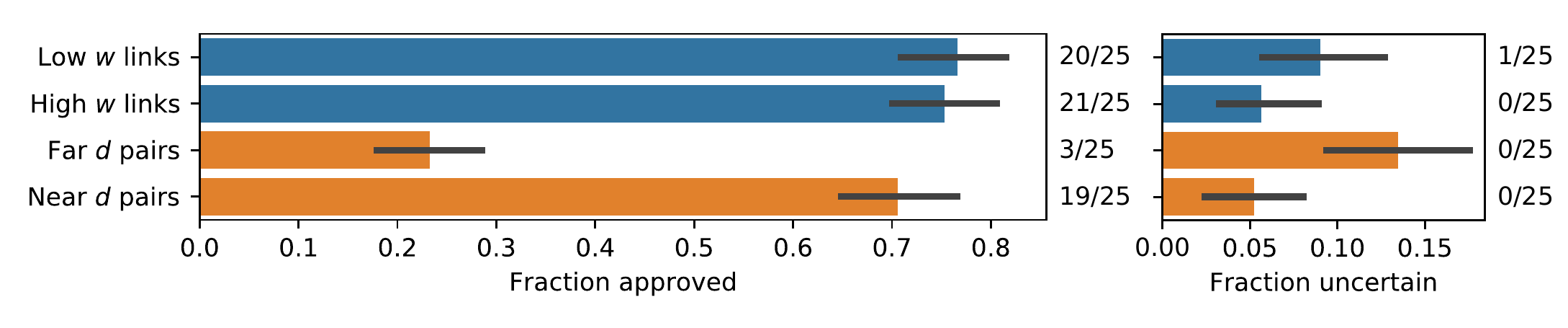}}
\caption{A followup survey asking workers who did not participate in Experiment 2 if they agree, disagree or were uncertain about cause-effect links and unlinked potential cause-effect pairs taken from the network generated by Experiment 2. Errorbars denote 95\% C.I.
Plotted bars denote the response fractions pooled across all surveyed links or pairs; numbers to the right denote counts of links or pairs where the \emph{majority} of responses to that individual link or pair agreed.
\label{fig:relQualSurvey}
}
\end{figure}

The results of these followup surveys are summarized in Fig.~\ref{fig:relQualSurvey} where we present the fraction of workers who approved or were uncertain about a potential cause-effect pair.
Most workers tended to approve of links already present in the network, even the low weight links (although workers were more uncertain about low-weight links).
Workers also approved causal attributions for most of the near distance ($d=2$), unlinked pairs but were significantly less likely to approve cause-effect pairs at high distance 
(z-test on proportions: $z = -10.5, p < 10^{-25}$) 
(and workers were significantly more likely to be uncertain about high distance pairs; $z = 3.14, p < 0.002$).
While further study is needed, these survey results do fit with our expectations if the data generated during Experiment 2 were of high quality, providing some evidence that the efficiency gains from our algorithms were not offset by lower quality or at least lower consistency, as perceived by other crowd workers.

\section{Computational study --- bias and robustness of network exploration}
\label{sec:model}

In this section we explore the differences between a true network and one derived from a sampling procedure based on Iterative Pathway Refinement (IPR).
We introduce a simplified computational model to simulate workers generating pathways and apply this model within the IPR algorithm on a known network topology to test IPR's performance.
As such, there are three components to this computational study: 
(i) a model of the underlying graph being explored (Sec.~\ref{subsec:modelGtrue}),
(ii) a model for how crowd workers develop pathways taken from the underlying graph (Sec.~\ref{subsec:modelingIPR}),
and (iii) a set of network metrics to compare the true underlying graph with the sampled graph derived from a finite number of worker pathways (Sec.~\ref{subsec:networkmetrics}).
We describe the results of our computational study using these components in Sec.~\ref{subsec:compstudyresults}.

\subsection{Modeling the underlying network}
\label{subsec:modelGtrue}

We study two directed network models for the underlying graph $\Gtrue$:

\begin{description}
\item[Random graph]
The directed analog of the undirected Erd{\H o}s-R\'enyi (ER)  graph~\cite{erdos1959random}.
Here $N$ nodes are introduced and each possible edge $(i,j)$ exists independently with constant probability $p$.
To make this network directed, both edge $(i,j)$ and $(j,i)$ exist with probability $p$, giving a maximum of $2\binom{N}{2}$ possible edges.

\item[Scale-Free graph]
The most common model for a scale-free graph is the  Barab\'asi-Albert (BA) graph~\cite{barabasi1999emergence}. Here
$N$ nodes are introduced one at a time starting from a small seed graph. Each new node
attaches to $m$ existing nodes with probability proportional to the existing nodes' degrees.
This leads to a rich-get-richer feedback mechanism as nodes with higher degree receive more incoming links giving them higher degree still.

Traditionally, the BA model is taken as an undirected network. The most obvious way to form a directed graph is where each newly introduced node $i$ forms a directed link $(i,j)$ when it attaches to a preexisting node $j$.
Unfortunately, this creates a ``temporal'' hierarchy of newer nodes pointing backwards to older nodes, and makes it impossible for any local search algorithm to explore the entire network.
To address this, we instead model directed edges as being equally likely to point in either direction, i.e., when a new node $i$ attaches to a preexisting node $j$, directed edge $(i,j)$ is created with probability 1/2, otherwise directed edge $(j,i)$ is added to the graph. 

\end{description}
We took $N = 1000$, $p=5/999$, and $m=3$ to simulate these networks, and 
only consider the giant connected component of $\Gtrue$ if it is disconnected.

The choice of these two model networks is meant to cover the extremes of possible degree distributions of the underlying network $\Gtrue$. If there is a strong dependence between the degree distribution and the efficiency of exploration, then it is likely that custom exploration algorithms will be needed for different networks. Conversely, however, if there is at most a weak dependence, then that means the exploration algorithm we propose may be able to efficiently explore a general class of networks.

\subsection{Modeling pathway refinement}
\label{subsec:modelingIPR}

Given a true, latent network to be explored, we seek a simple model for how workers could reveal that network structure via individual IPR microtasks.
Network search models, where an agent moves locally over a network exploring its structure, provide inspiration. 
Here we treat the pathway refinement problem as one of a short-term local search of a network. 
A searcher, starting from a randomly chosen node in $\Gsampled$ and respecting edge directions, takes a small number of moves to connected nodes, either in $\Gsampled$ or in $\Gtrue$.
The trace of the searcher can then mimic a new pathway generated by a worker proposing existing or novel terms.

Specifically, we model pathway generation using \textbf{self-avoiding walks} (SAW).
A SAW on a graph is a variant of a random walk where the random walker is constrained such that it cannot visit the same node more than once~\cite{de1979scaling,li1995critical,herrero2005self}. 
Of course, independent SAWs my overlap, traversing some of the same nodes and links as other SAWs.
Short-length SAWs can then represent causal pathways that do not repeat terms---almost no pathways generated during Experiment 2 contained duplicate terms (Sec.~\ref{sec:exp2}).
Self-avoiding walks have good properties for efficiently exploring unknown networks~\cite{adamic2001search,PhysRevE.71.016107}, providing a theoretical underpinning for the IPR algorithm.

To simulate IPR exploring a latent network with this model, 
at each timestep a new pathway is generated by an independent self-avoiding walker initialized at a node within $\Gsampled$.
This walk proceeds for a number of steps $L$, and any new nodes or links the walker happens to travel over---those in $\Gtrue$ not yet in $\Gsampled$---are added to $\Gsampled$.
In our simulations, the length of each pathway $L$ was drawn from a shifted Poisson distribution with fixed mean $\lambda = 1$, i.e.\ $L \sim \mathrm{Pois}(\lambda=1)+ \ell$. 
We selected constant values of $\ell = 4,6, 8$ that bracket the mean path length observed in Experiment 2 (see Fig. ~\ref{fig:pathwaySatisficing}).
The stochasticity introduced by this distribution is intended to better mimic the variation in length of actual crowdsourced pathways, although the exact form of the statistical model is not crucial.
Further, choosing (constant) $L = 1$ corresponds to the Single Link task studied in Experiment 1 (Sec.~\ref{sec:exp1}), so we can compare simulated IPR exploration with Single Link exploration.
This model can thus build $\Gsampled$ out of $\Gtrue$ over time by taking the union of all the simple model pathways created, providing a plausible, though idealized, simulation of the Iterative Pathway Refinement crowdsourcing phase.

\subsection{Network exploration metrics}
\label{subsec:networkmetrics}

Suppose $\Gsampled$ is derived from simulating our IPR model (Sec.~\ref{subsec:modelingIPR}) until a total of $\left|P\right|$ pathways have been developed.
To understand the effects of generating a network via this algorithm requires comparing $\Gsampled$ with the true, latent network $\Gtrue$. 
Of course, in practice $\Gtrue$ is unavailable for such comparisons, but the synthetic analysis performed here helps us understand, in some manner, the performance and properties of our crowdsourcing algorithm.

To compare $\Gsampled$ and $\Gtrue$ we use the following network metrics:
\begin{description}

\item[Fraction of nodes and links observed]
Here we track how much of the underlying network $\Gtrue$ is captured in $\Gsampled$ by comparing the numbers of nodes $N$ and links $M$ in the observed graph with the sizes of the true graph: $\Nsampled / \Ntrue$ and $\Msampled /\Mtrue$ as functions of $\left|P\right|$. 

\item[Average degree]
We also track the mean indegree and outdegree $\left< k \right>$ of nodes in $\Gsampled$ compared with the same quantities for $\Gtrue$. 
In the context of causal attribution networks, indegree captures the number of causes an effect has, while outdegree captures the number of effects that are due to a cause.
This metric captures what portion of the neighbors of nodes are retained by the sampling algorithm.

\item[Betweenness centrality]
The betweenness centrality $B(v)$ of a node $v$ in $G$ is the sum of the proportion of all shortest paths that pass through $v$~\cite{brandes2008variants}:
\begin{equation}
B(v; G) =\sum_{s,t \in V} \frac{\sigma(s, t|v)}{\sigma(s, t)}
\label{eqn:btwn}
\end{equation}
where $\sigma(s, t)$ is the number of
shortest paths connecting nodes $s$ and $t$, and $\sigma(s, t | v)$ is the number of
those paths passing through another node $v$ ($v \neq s$, $v \neq t$).
To compare $\Gsampled$ and $\Gtrue$, we take the ratio $\Bsampled / \Btrue$ using the mean node betweennesses $\Bsampled = \frac{1}{\Nsampled} \sum_{v \in \Vsampled} B(v;\Gsampled)$ and $\Btrue = \frac{1}{\Ntrue} \sum_{v \in \Vtrue} B(v;\Gtrue)$.

\end{description}

The focus here is on relatively basic metrics that quantify global differences in the networks, but many other network metrics and statistics are available.
For example, one statistic commonly used to quantify random walkers searching a network, particularly in the statistical physics literature, is the mean first-passage time, measuring how long it takes an individual random walker to reach a particular node~\cite{redner2001guide,PhysRevE.71.016107}. 
Yet, this measure is most suitable for a single random walk that moves over the network, whereas our model focuses on the collective action of a large number of shorter walks, so we instead focus on metrics suitable to quantifying the proportion of the true network that has been revealed, and if and how properties of $\Gsampled$ differ from those of $\Gtrue$.

\subsection{Results}
\label{subsec:compstudyresults}

\begin{figure}
\centering
{\includegraphics[width=0.9\textwidth,trim=0 12 0 8,clip=true]{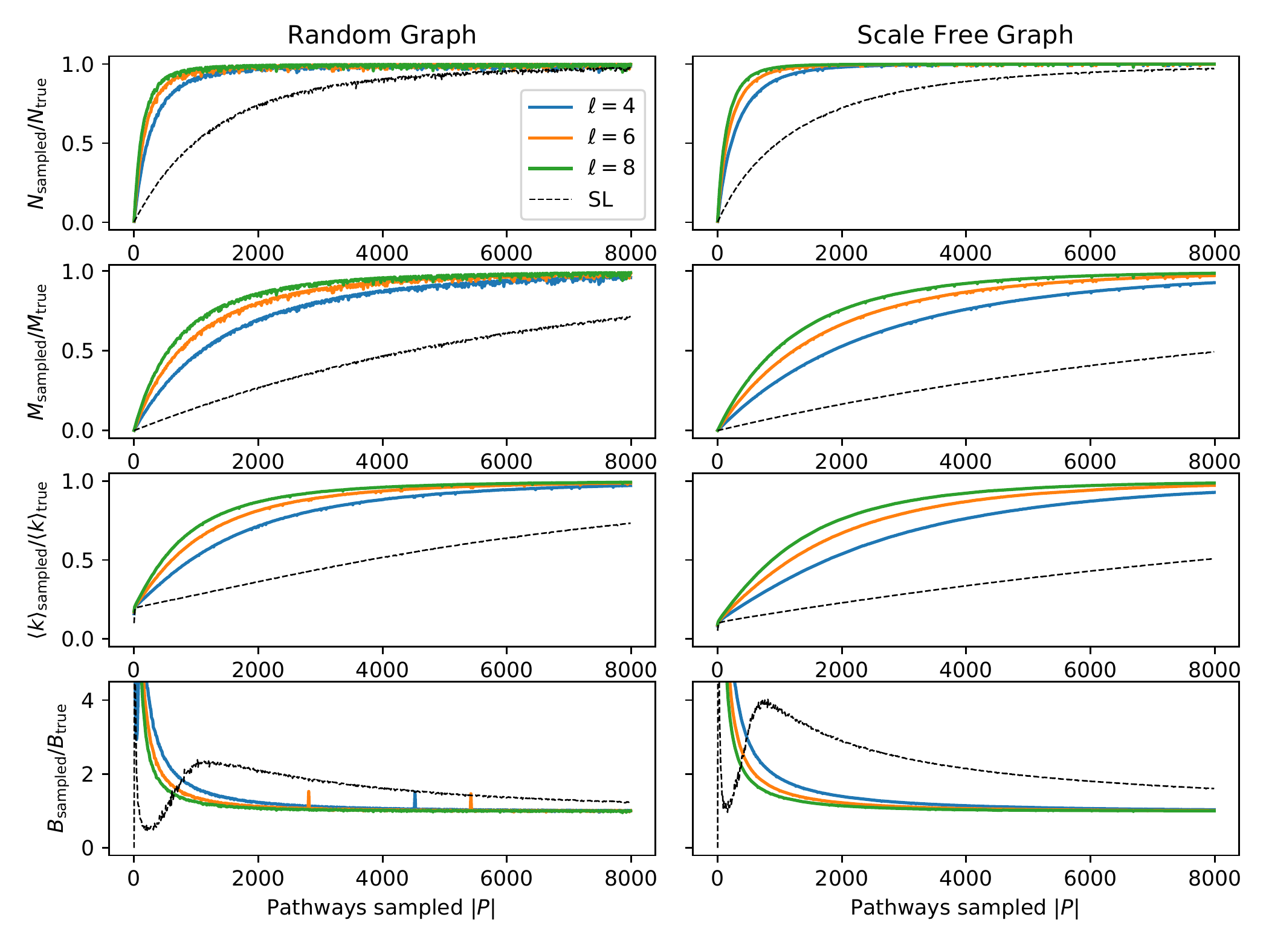}}
\caption{Convergence of $\Gsampled$ to $\Gtrue$ using the self-avoiding walk (SAW) model of IPR. 
The Single Link (SL) learning baseline represents the task studied in Experiment 1 (Sec.~\ref{sec:exp1}).
\label{fig:sampledVsTrue}
}
\end{figure}

Figure \ref{fig:sampledVsTrue} presents the results of our computational study (simulating 100 independent runs for each combination of parameters).
For both the random and scale-free networks, IPR was able to explore the latent $\Gtrue$ more quickly than the single link task:
The fraction of nodes discovered, fraction of links discovered, and average node degree all converged more quickly to their true values, especially for longer average $L$.
For example, IPR with $\ell = 4$ revealed 90\% of the nodes in the Scale-Free graph on average with only $\approx 940$ IPR tasks while over $4200$ single link tasks were necessary to reach the same proportion, a factor of $4.6$ more tasks.
Of course, longer pathways should necessarily reveal the network more quickly, because each pathway provides more subnetwork structure, but in practice this will be complemented by the faster speed of IPR workers per link observed in Sec.~\ref{subsec:speedIPRworkers}.
Overall, IPR is effective at exploring different types of networks.

Surprisingly, unlike the other metrics, the mean betweenness for $\Gsampled$ is generically \emph{higher} than it is for $\Gtrue$.
This reveals an important bias: short-duration explorations will over-report the centralities of nodes, even though betweenness centrality (Eq.~\eqref{eqn:btwn}) already accounts for the total number of shortest paths in $\Gsampled$.
The origin of this bias is that there are fewer links early on in the exploration, leading to fewer alternate shortest paths and so more paths will ``flow'' through the currently observed shortest paths.
As more nodes and links are added, new shortest paths will appear, taking away some of the load carried by the existing shortest paths.
Thus, a crowdsourcer exploring a network should account for biases in betweenness centrality and possibly other centrality measures.
This biased overestimation of betweenness at low numbers of samples is far stronger for the single-link sampling method than for the IPR model, further emphasizing the usefulness of the IPR method.

\section{Discussion and future work}
\label{sec:discussion}

Establishing causal relationships is one of the most important and challenging tasks of science.
This work focused on crowdsourcing causal attribution networks, directed networks where links $A \to B$
denote that a term $A$ is a cause and the term $B$ is its effect.
However, much of the methodology contributed in this work can inform the exploration of other types of networks, 
for example, surveying individuals and their social ties to map out a social network~\cite{heckathorn1997respondent}.
Many technical networks can be explored if the participants possess the relevant expertise. A team of cardiology researchers and bioinformaticians, for instance, could collaboratively build a network of gene pathways related to cardiac arrest.
Such domain-specific cases are an especially interesting avenue to combine our network exploration method with statistical methods and data, such as gene expression datasets for the cardiology example.

The crowdsourced causal attribution network derived by this study warrants further refinement, by merging synonymous terms, filtering out unsupported links, ``backfilling'' incomplete portions of the network as needed, and otherwise accounting for the human perceptual biases that influence causal reasoning.
Indeed, while our first experiment provided evidence that workers are generally able to recognize true causal relationships, workers also make false positives, attributing cause and effect relationships where none  exist.
Measuring and accounting for this bias remains an important challenge when developing larger causal networks.

Our multiphase network crowdsourcing algorithm enables \emph{\textbf{indirect collaboration}} between workers, as workers only interact with the responses given by previous workers when viewing those responses as part of a subsequently constructed task.
Direct collaboration, where teams of individuals work together to build causal pathways, is an interesting area we plan to explore. 
This approach may lead to better causal attribution data, novel information on how teams perform, and improved understanding of the perceptual and attributional biases of causal relationships.

Iterative Pathway Refinement (IPR) is the core of our crowdsourcing algorithm. 
Our experiments, followup survey, and computational study provide evidence that IPR can leverage crowdsourcing to explore a large network relatively efficiently.
IPR enabled workers to provide information efficiently and is well motivated by prior theoretical work on exploring networks using self-avoiding walks.
Workers complete IPR tasks by modifying short linear causal pathways or chains. 
IPR is a special case of the more general \emph{\textbf{Iterative Motif Refinement}} (IMR), as these linear chains can easily generalize to other types of motifs. 
Chains were chosen for IPR as they are relatively easy to understand in the context of cause-and-effect, but other network motifs can be used as well.

Indeed, studying motifs as part of IMR is a particularly fruitful line of future research, as different motifs may work best when exploring different types of networks, particularly when accounting for differences in human judgment and perception across the different types of networks.
For example, a chain (dipath) may be most appropriate for a causal graph, while a triangle motif may be better when crowdsourcing a social network, as triadic closure is one of the most important properties of social networks~\cite{Rapoport1953}.
If training data are available, one could even in principle use supervised learning to determine the ideal motif for a given network automatically.

Iterative Pathway Refinement can also be improved in other ways.
We focused on one of the \emph{simplest possible forms} of IPR, where existing pathways were chosen uniformly at random to show to workers.
Although the same pathway was prevented from being shown too frequently (by limiting the total number of responses to a given pathway),
IPR would be more efficient still if existing pathways were selected for updating by incorporating our current understanding of the quality or correctness of the given pathway. 
This would better distribute the crowd by preventing repeated refinement of known good pathways and instead guide the crowd towards those pathways that are the most fruitful to explore.
Likewise, network-specific algorithms such as IPR can be complemented by non-network crowdsourcing algorithms such as efficient response aggregation strategies~\cite{dawid1979maximum,karger2011iterative,kruger2014axiomatic}.
Algorithms that balance an exploration-exploitation trade-off, such as Thompson sampling~\cite{NIPS2011_4321}, are also likely to improve the exploration efficiency of IPR even further.

\begin{acks}
We thank the anonymous reviewers for their helpful comments and the crowd workers whose participation made this research possible.
This material is based upon work supported by the National Science Foundation under Grant No.\ IIS-1447634.
\end{acks}

\bibliographystyle{ACM-Reference-Format}
\bibliography{main}

\appendix
\section[]{Mechanical Turk Worker Instructions}
\label{app:mturkInstructions}

\begin{figure}[h]
\centering
\frame{\includegraphics[width=0.55\textwidth]{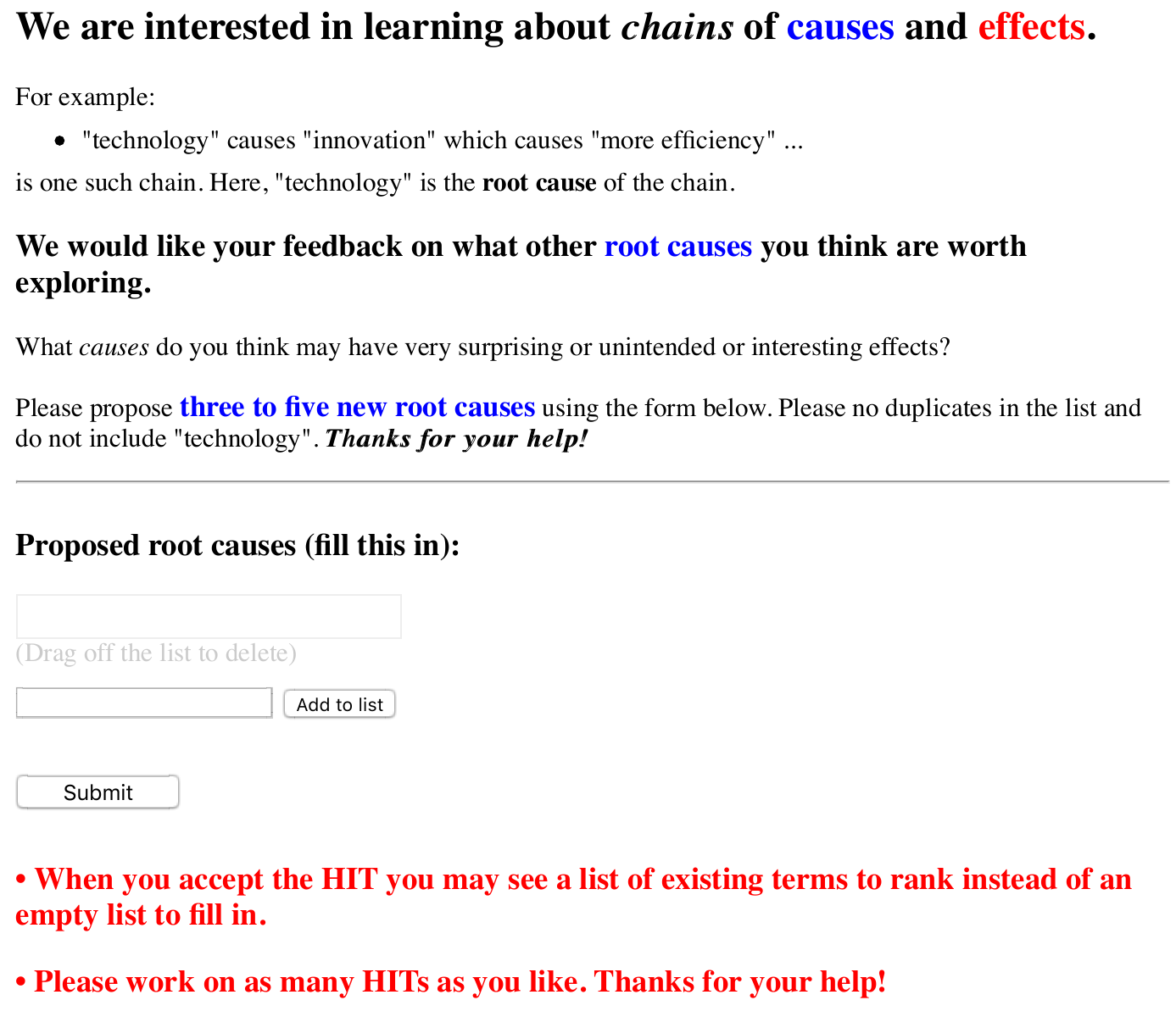}}
\caption{Screenshot of the instructional page shown to Amazon Mechanical Turk workers before accepting the Cause Proposal task that initialized Experiment 2.
\label{fig:screenshotCPinstructions}
}
\end{figure}

\begin{figure}[h]
\centering
  \subcaptionbox{Greedy Pathway Expansion\label{third-subfig}}{%
	\frame{\includegraphics[width=0.49\textwidth,trim=20 130 155 113,clip=true]{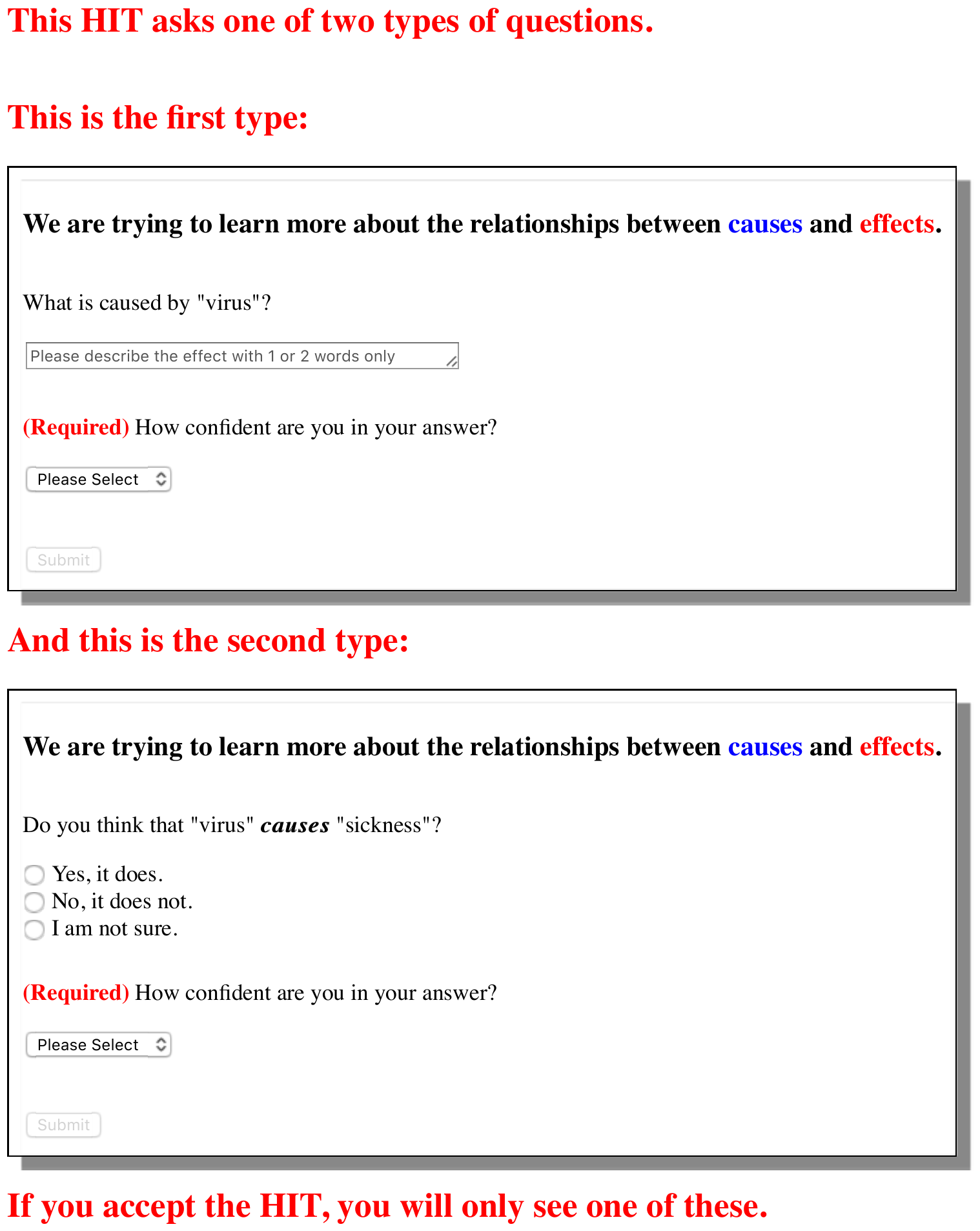}}%
  }
  \subcaptionbox{Iterative Pathway Refinement (cf.\ Fig.~\ref{fig:IPRformscreenshot}) \label{fourth-subfig}}{%
{\includegraphics[width=0.49\textwidth,trim=20 370 245 235,clip=true]{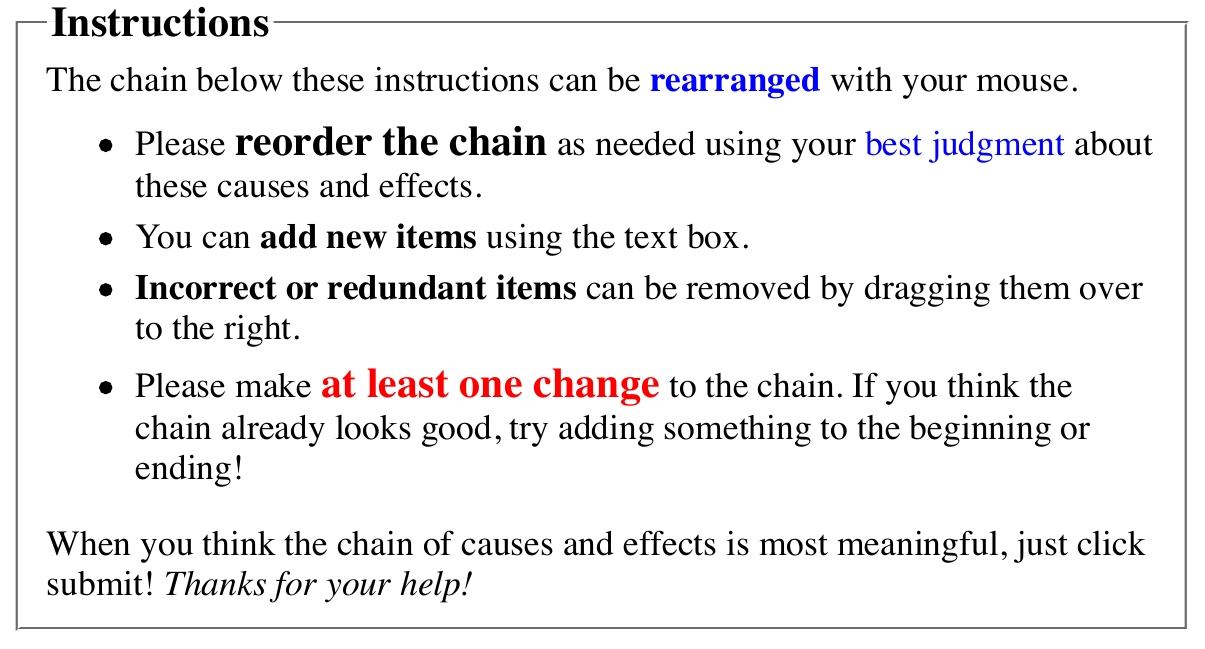}}%
  }
\caption{Screenshots of the instructional pages shown to Amazon Mechanical Turk workers before accepting (a) Greedy Pathway Expansion tasks and (b) Iterative Pathway Refinement tasks. 
For GPE tasks, a worker will be asked to either propose a new effect or to validate an existing cause-effect  pair, depending on the state of the algorithm, and so worker instructions cover both tasks.
\label{fig:screenshotGPEinstructions}
}
\end{figure}

Here are the instructional pages AMT workers were shown before they accepted each type of HIT comprising Experiment 2. 
Figure \ref{fig:screenshotCPinstructions} shows the instructional page for workers asked to propose or rank seed causes, Fig.~\ref{fig:screenshotGPEinstructions}(a) shows the instructional page for workers participating in the Greedy Pathway Expansion phase, and Fig.~\ref{fig:screenshotGPEinstructions}(b) shows the instructional page for workers participating in the Iterative Pathway Refinement phase. 
Figure~\ref{fig:screenshotGPEinstructions}(b) shows the contents of the collapsed instructions box found in Fig.~\ref{fig:IPRformscreenshot}; the box was expanded for the instructional page but collapsed by default during actual HITs.
The ``vote'' phase shown at the bottom of Fig.~\ref{fig:screenshotGPEinstructions}(a) is also the task interface used for the response quality survey (Sec.~\ref{sec:followupsurvey}).


\end{document}